\documentclass[12pt, onecolumn,draftclsnofoot]{IEEEtran}%

\usepackage[T1]{fontenc}
\usepackage{amsmath,amssymb,amsfonts,mathrsfs,bm}
\usepackage{mathtools}
\usepackage{amsthm}
\usepackage{nicefrac}
\usepackage{cite}
\usepackage[shortlabels]{enumitem}
\usepackage{graphicx}
\usepackage{epstopdf}
\usepackage{url}
\usepackage{colortbl}
\usepackage{booktabs}
\usepackage{multirow}
\usepackage[table,dvipsnames]{xcolor}
\usepackage[normalem]{ulem}

\usepackage{array}
\newcolumntype{L}[1]{>{\raggedright\let\newline\\\arraybackslash\hspace{0pt}}m{#1}}
\newcolumntype{C}[1]{>{\centering\let\newline\\\arraybackslash\hspace{0pt}}m{#1}}
\newcolumntype{R}[1]{>{\raggedleft\let\newline\\\arraybackslash\hspace{0pt}}m{#1}}

\makeatletter
\let\MYcaption\@makecaption
\makeatother
\usepackage[font=footnotesize]{subcaption}
\makeatletter
\let\@makecaption\MYcaption
\makeatother

\usepackage{xparse}



\usepackage{glossaries}

\makeatletter
\let\oldgls\gls
\let\oldglspl\glspl

\newcommand\fussy@ifnextchar[3]{%
  \let\reserved@d=#1%
  \def\reserved@a{#2}%
  \def\reserved@b{#3}%
  \futurelet\@let@token\fussy@ifnch}
\def\fussy@ifnch{%
  \ifx\@let@token\reserved@d
    \let\reserved@c\reserved@a 
  \else
    \let\reserved@c\reserved@b
  \fi
 \reserved@c}

\renewcommand{\gls}[1]{%
  \oldgls{#1}\fussy@ifnextchar.{\@checkperiod}{\@}}
\renewcommand{\glspl}[1]{%
  \oldglspl{#1}\fussy@ifnextchar.{\@checkperiod}{\@}}

\newcommand{\@checkperiod}[1]{%
  \ifnum\sfcode`\.=\spacefactor\else#1\fi
}
\makeatother

\newacronym{wrt}{w.r.t.}{with respect to}
\newacronym{RHS}{RHS}{right-hand side}
\newacronym{LHS}{LHS}{left-hand side}
\newacronym{iid}{i.i.d.}{independent and identically distributed}

\usepackage{float}

\ifx\notloadhyperref\undefined
	\ifx\loadbibentry\undefined
		\usepackage[hidelinks,hypertexnames=false]{hyperref} 
	\else
		\usepackage{bibentry}
		\makeatletter\let\saved@bibitem\@bibitem\makeatother
		\usepackage[hidelinks,hypertexnames=false]{hyperref}
		\makeatletter\let\@bibitem\saved@bibitem\makeatother
	\fi
\else
	\ifx\loadbibentry\undefined
		\relax
	\else
		\usepackage{bibentry}
	\fi
\fi

\usepackage[capitalize]{cleveref}
\crefname{equation}{}{}
\Crefname{equation}{}{}
\crefname{claim}{claim}{claims}
\crefname{step}{step}{steps}
\crefname{line}{line}{lines}
\crefname{condition}{condition}{conditions}
\crefname{dmath}{}{}
\crefname{dseries}{}{}
\crefname{dgroup}{}{}

\crefname{Problem}{Problem}{Problems}
\crefformat{Problem}{Problem~(#2#1#3)}
\crefrangeformat{Problem}{Problems~(#3#1#4) to~(#5#2#6)}

\crefname{Theorem}{Theorem}{Theorems}
\crefname{Corollary}{Corollary}{Corollaries}
\crefname{Proposition}{Proposition}{Propositions}
\crefname{Lemma}{Lemma}{Lemmas}
\crefname{Definition}{Definition}{Definitions}
\crefname{Example}{Example}{Examples}
\crefname{Assumption}{Assumption}{Assumptions}
\crefname{Remark}{Remark}{Remarks}
\crefname{Rem}{Remark}{Remarks}
\crefname{remarks}{Remarks}{Remarks}
\crefname{Appendix}{Appendix}{Appendices}
\crefname{Exercise}{Exercise}{Exercises}
\crefname{Theorem_A}{Theorem}{Theorems}
\crefname{Corollary_A}{Corollary}{Corollaries}
\crefname{Proposition_A}{Proposition}{Propositions}
\crefname{Lemma_A}{Lemma}{Lemmas}
\crefname{Definition_A}{Definition}{Definitions}

\usepackage{crossreftools}
\ifx\notloadhyperref\undefined
\else
	\relax
\fi

\usepackage{algorithm,algorithmic}

\ifx\loadbreqn\undefined
	\relax
\else
	\usepackage{breqn} 
\fi

\ifx\renewtheorem\undefined
\ifx\useTheoremCounter\undefined
\newtheorem{Theorem}{Theorem}
\newtheorem{Corollary}{Corollary}
\newtheorem{Proposition}{Proposition}
\newtheorem{Lemma}{Lemma}
\else
\newtheorem{Theorem}{Theorem}

\newtheorem{Proposition}[theorem]{Proposition}
\fi

\newtheorem{Definition}{Definition}
\newtheorem{Example}{Example}

\fi

\theoremstyle{remark}

\theoremstyle{plain}

\DeclareSymbolFont{bsfletters}{OT1}{cmss}{bx}{n}
\DeclareSymbolFont{ssfletters}{OT1}{cmss}{m}{n}
\DeclareMathSymbol{\bsfGamma}{0}{bsfletters}{'000}
\DeclareMathSymbol{\ssfGamma}{0}{ssfletters}{'000}
\DeclareMathSymbol{\bsfDelta}{0}{bsfletters}{'001}
\DeclareMathSymbol{\ssfDelta}{0}{ssfletters}{'001}
\DeclareMathSymbol{\bsfTheta}{0}{bsfletters}{'002}
\DeclareMathSymbol{\ssfTheta}{0}{ssfletters}{'002}
\DeclareMathSymbol{\bsfLambda}{0}{bsfletters}{'003}
\DeclareMathSymbol{\ssfLambda}{0}{ssfletters}{'003}
\DeclareMathSymbol{\bsfXi}{0}{bsfletters}{'004}
\DeclareMathSymbol{\ssfXi}{0}{ssfletters}{'004}
\DeclareMathSymbol{\bsfPi}{0}{bsfletters}{'005}
\DeclareMathSymbol{\ssfPi}{0}{ssfletters}{'005}
\DeclareMathSymbol{\bsfSigma}{0}{bsfletters}{'006}
\DeclareMathSymbol{\ssfSigma}{0}{ssfletters}{'006}
\DeclareMathSymbol{\bsfUpsilon}{0}{bsfletters}{'007}
\DeclareMathSymbol{\ssfUpsilon}{0}{ssfletters}{'007}
\DeclareMathSymbol{\bsfPhi}{0}{bsfletters}{'010}
\DeclareMathSymbol{\ssfPhi}{0}{ssfletters}{'010}
\DeclareMathSymbol{\bsfPsi}{0}{bsfletters}{'011}
\DeclareMathSymbol{\ssfPsi}{0}{ssfletters}{'011}
\DeclareMathSymbol{\bsfOmega}{0}{bsfletters}{'012}
\DeclareMathSymbol{\ssfOmega}{0}{ssfletters}{'012}

\newcommand{\balpha}{\bm{\alpha}}

\DeclareMathOperator*{\argmin}{arg\,min}

\newcommand{\qednew}{\nobreak \ifvmode \relax \else
      \ifdim\lastskip<1.5em \hskip-\lastskip
      \hskip1.5em plus0em minus0.5em \fi \nobreak
      \vrule height0.75em width0.5em depth0.25em\fi}

\newcommand{\norm}[1]{{\left\lVert{#1}\right\rVert}}

\DeclareDocumentCommand\set{s m t| m}{%
  \IfBooleanTF#1%
	{\left\{\, #2\mathrel{} \IfBooleanTF{#3}{\middle|}{:}\mathrel{}  #4\, \right\}}%
  {\{\, #2 \IfBooleanTF{#3}{\mid}{\mathrel{} : \mathrel{}} #4\, \}}%
}

\DeclareDocumentCommand \ifcond {m m} {%
	{#1} %
	\IfValueT{#2}{\, \middle|\, {#2}}%
}

\DeclareDocumentCommand \P {e{_} g >{\SplitArgument{ 1 }{ @| }}d() g } {%
	\mathbb{P}%
	\IfValueTF{#1}{_{#1}}
		{\IfValueT{#2}{_{#2}}}%
	\IfValueT{#3}{\left(\ifcond#3}%
	\IfValueT{#4}{\, \middle|\, {#4}}%
	\IfValueT{#3}{\right)}%
}

\DeclareDocumentCommand \E {e{_} g >{\SplitArgument{ 1 }{ @| }}o g } {%
	\mathbb{E}%
	\IfValueTF{#1}{_{#1}}
		{\IfValueT{#2}{_{#2}}}%
	\IfValueT{#3}{\left[\ifcond#3}%
	\IfValueT{#4}{\, \middle|\, {#4}}%
	\IfValueT{#3}{\right]}%
}

\definecolor{gray90}{gray}{0.9}

\ifx\nohighlights\undefined

	\newcommand{\msout}[1]{\text{\color{green} \sout{\ensuremath{#1}}}}
	\newcommand{\del}[1]{{\color{green}\ifmmode \msout{#1}\else\sout{#1}\fi}}
\else

	\newcommand{\msout}[1]{#1}
	\newcommand{\del}[1]{#1}
\fi

\newcommand{\hhide}[1]{}

\renewcommand{\figurename}{Fig.}
\newcommand{\figref}[1]{\figurename~\ref{#1}}
\graphicspath{{./Figures/}} 
\newcommand{\includeCroppedPdf}[2][]{%
    \IfFileExists{./Figures/#2-crop.pdf}{}{%
        \immediate\write18{pdfcrop ./Figures/#2 ./Figures/#2-crop.pdf}}%
    \includegraphics[#1]{./Figures/#2-crop.pdf}}

\ifx\diagnoselabel\undefined
	\relax
\else
	\makeatletter
	 \def\@testdef #1#2#3{%
		 \def\reserved@a{#3}\expandafter \ifx \csname #1@#2\endcsname
		\reserved@a  \else
	 \typeout{^^Jlabel #2 changed:^^J%
	 \meaning\reserved@a^^J%
	 \expandafter\meaning\csname #1@#2\endcsname^^J}%
	 \@tempswatrue \fi}
	\makeatother
\fi

\crefname{question}{question}{questions}

\usepackage{tikz}
\usepackage{pgfplots}
\pgfplotsset{compat=1.5}

\providecommand{\U}[1]{\protect\rule{.1in}{.1in}}

\theoremstyle{definition}

\crefname{prob}{Problem}{Problems}

\begin{document}
\date{}
\title{To further understand graph signals}
\author{Feng~Ji, Wee~Peng~Tay,~\IEEEmembership{Senior Member,~IEEE} 
\thanks{The authors are with the School of Electrical and Electronic Engineering, Nanyang Technological University, 639798, Singapore (e-mail: jifeng@ntu.edu.sg, wptay@ntu.edu.sg.)}%
}
\maketitle
\begin{abstract}
Graph signal processing (GSP) is a framework to analyze and process graph-structured data. Many research works focus on developing tools such as Graph Fourier transforms (GFT), filters, and neural network models to handle graph signals. Such approaches have successfully taken care of ``signal processing'' in many circumstances. In this paper, we want to put emphasis on ``graph signals'' themselves. Although there are characterizations of graph signals using the notion of bandwidth derived from GFT, we want to argue here that graph signals may contain hidden geometric information of the network, independent of (graph) Fourier theories. We shall provide a framework to understand such information, and demonstrate how new knowledge on ``graph signals'' can help with ``signal processing''.   
\end{abstract}


\section{Introduction} \label{sec:intro}
Since its emergence, the theory and applications of graph signal processing (GSP) have rapidly developed \cite{Shu12, Shu13, San13, San14, Gad14, Don16, Def16, Kip16, Egi17, Gra18, Ort18, Girault2018, Ji19}. Many different aspects of GSP have been explored. GSP is based on the choice of a graph shift operator (GSO), and the cornerstone of GSP is \emph{graph Fourier transform} defined using the GSO \cite{Shu12, Shu13}. This allows us to introduce \emph{frequency domain}, analogous to the classical Fourier theory. A well-studied topic in GSP is the \emph{theory of filtering} \cite{Shu13, Ort18}. Graph filters are fundamental tools to analyze and process graph signals. Many important topics stem from the theory of filtering, including \emph{sampling theory} \cite{Aga13, Che15, Tsi15, Mar16, Anis2016} and \emph{graph neural networks} \cite{Def16, Kip16}. The article \cite{Ort18} contains a comprehensive overview that also discusses many other related topics such as signal reconstruction, graph learning, and applications of GSP. 

Suppose a graph $G$ is of size $n$. According to the definition, a graph signal $f$ is a vector in $\mathbb{R}^n$, with each component corresponding to a node of the graph. There is no reasonable interpretation of $f$ without the graph $G$. The above-mentioned works take care of the ``signal processing'' aspect of GSP, by leveraging the fundamental assumption that the signal value should be close to each other at any pair of nodes connected by an edge. From here, we see that the property of the signal $f$ with respective to (w.r.t.) the graph $G$ plays a central role. Motivated by such a consideration, in this paper, we focus solely on the ``graph signal'' aspect of GSP. To be more precise, we want to describe how we may quantify the notion of ``smoothness'' of graph signals. Based on such a notion, we can give a characterization of graph signals. To do so, we want to dig out hidden geometric information contained in graph signals and compare such information with the geometry of the given graph. 

Classically, many research works rely on the notion of bandlimitedness \cite{Anis2016, Tza18, Roh19, Bar19, Jif20} to characterize graph signals, with tasks ranging from sampling, and signal reconstruction to topology learning. A signal with small bandwidths is considered to be smooth. However, it is arguable whether such consideration is most appropriate or not. For example, in the first place, bandlimitedness does not only depend on the graph but also relies heavily on the choice of the GSO, for which we have quite a few candidates. Moreover, for most of the common choices of GSO, smooth signals are those whose values are close to each other at any pair of nodes connected by an edge, as we described earlier. However, signal values between nodes further away are not directly compared. The approach taken in our paper shall address such shortcomings.

To set a few humble goals, we want to:
\begin{itemize}
    \item Develop new methods to identify hidden geometric information of graph signals.
    \item Use these methods to explicitly define the smoothness of graph signals.
    \item Classify graph signals based on the new notion of smoothness.
    \item Investigate new insights into GSP by combining a new understanding of graph signals with classical GSP tools.  
\end{itemize}

Each of these goals is fully explored in a section of the paper, and the rest of the paper is organized as follows. We motivate our goals and approach in \cref{sec:gst}. We realize that asides from being either smooth or noisy, a graph signal can have other characterizations that agree with neither. For example, as we demonstrate with an example, a signal can look noisy but contains important geometric information. We coin the term ``perpendicular signal'' for such a signal. To understand perpendicular signals, we propose to take an indirect route and study a quantitative notion of smoothness. In \cref{sec:cou}, we describe the framework to compare a graph and signals by using an explicit geometric construction. The framework us allows to formally define the smoothness of graph signals in \cref{sec:smo}. We present simulation results in \cref{sec:sim}. The focus is to combine our investigation of signals themselves with well-established GSP tools, to shed light on new insights into GSP theory. We finally conclude in \cref{sec:con}. All proofs are contained in the Appendix.

\section{Graph signal types} \label{sec:gst}

Let $G=(V,E)$ be a finite unweighted simple graph of size $n=|V|$, where $V$ is the vertex set and $E$ is the set of edges of $G$. A \emph{graph signal} $f$ is a function $f: V \to \mathbb{R}$. An equivalent interpretation is to view $f$ as an $n$-dimensional vector, where the component corresponding to $v\in V$ is denoted by $f(v)$. There have been numerous studies of graph signals with the theory of graph signal processing (GSP). One of the central themes is the study of \emph{bandlimited graph signals}. Intuitively, the notion is a discrete analog of its counterpart in classical Fourier theory. Such a signal is considered to be smooth in the sense that the signal values at neighboring nodes are close to each other. On the opposite side, we have \emph{noises}, whose signal values fluctuate widely even across neighboring nodes. There are statistical models for noises \cite{}. In general, they are regarded as obstacles in graph signal processing. In the next example, we want to discuss the possibility of scooping up useful information from ``noises''. The key observation is that a seemingly noisy graph signal may contain geometric information that supplements the graph structure. 

\begin{Example} \label{eg:cah}
Consider a helix curve (a spiral in $\mathbb{R}^3$ as in \figref{fig:11}) $C$ given by the parametric form 
\begin{align*} C(t) = (x(t),y(t),z(t))= (\cos t, \sin t, t), t\in [0,5\pi]. \end{align*} Assume that $n$ is chosen such that $5\nmid n-1$. We take $n$ uniformly spaced points $V = \{v_1,\ldots, v_n\} \subset C$ with $v_1=C(0)$ and $v_n = C(5\pi)$. The set $V$ together with their connections on $C$ gives rise to the path graph $P$ with $n$ nodes. On the other hand, we may equivalently encode all information in a different graph $G$ and a signal $f$ as follows. 

We apply the projection $p$ of $V$ to the $(x,y)$-plane. Based on their proximity on the unit circle, we have a cycle graph $G$ on $n$-nodes $u_i = p(v_i), 1\leq i\leq n$. The condition $5\nmid n-1$ ensures $u_i\neq u_j$ for $i\neq j$. We construct a graph signal $f$ on $G$ such that $f(u_i)$ is the height of $v_i$, i.e., $f(u_i) = 5(i-1)\pi/(n-1)$. As $C$ spirals multiple rounds, there are nodes close to each other on $G$ whose $f$ values differ much. Hence, $f$ resembles a noise. However, if we disregard the geometric information contained in $f$, we may have a wrong interpretation of both $G$ and $f$. In such cases, applying GSP tools to $f$ viewed as a signal on $G$ becomes inappropriate. However, $G$ and $f$ combined do contain full information about the original setup. Moreover, we observe that $f$ is placed in the direction \emph{perpendicular} to $G$. 
      \begin{figure}[!htb] 
	\centering
	\includegraphics[scale=0.7]{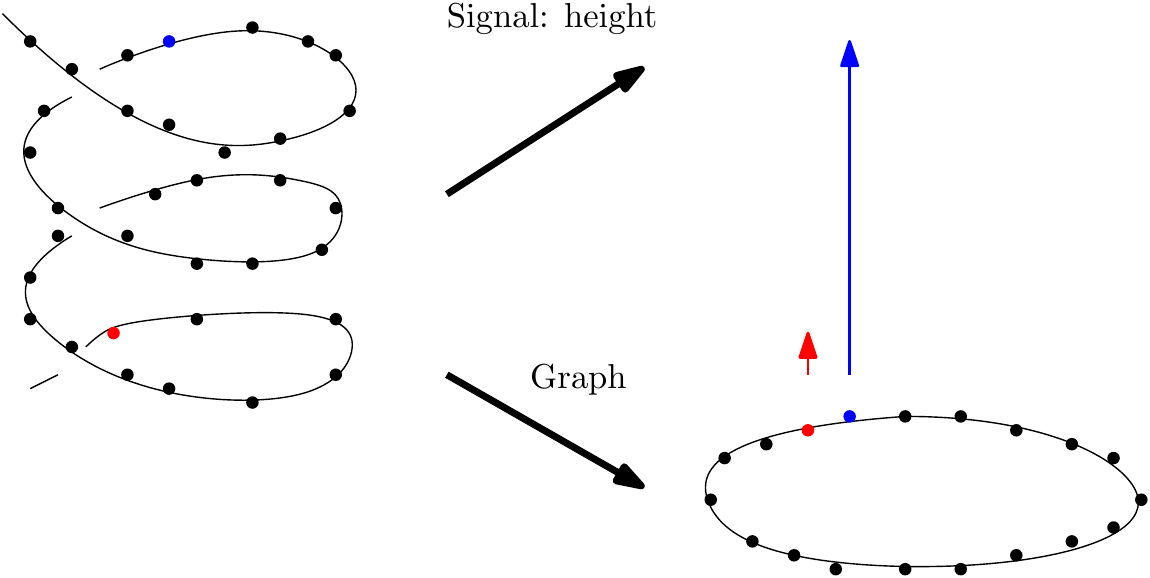}
	\caption{The helix curve.} \label{fig:11}
\end{figure}
\end{Example}

To give more intuitions, we discuss two different ways to interpret graph signals. Classically, one chooses a graph shift operator (GSO) $A$ and constructs a filter bank from it, such as the convolution filters, and band-pass filters. Given a graph $G=(V,E)$ and a signal $f$, one applies filters from a filter bank to $f$, for analysis purposes. This is the \emph{function point of view} of graph signals, namely we study properties of $f$ using the geometry of $G$ encoded in $A$. 

On the other hand, we may also consider the \emph{geometric point of view} of graph signals. Namely, we view $f$ as mapping the nodes $V$ into $\mathbb{R}$. This should provide additional geometric information of $V$ other than that contained in $E$. In particular, if such additional information is not coherent w.r.t.\ that of $E$ (e.g., \cref{eg:cah}), applying filters constructed from $G$ to analyze $f$ becomes inappropriate. 

The function point of view focuses on understanding graph signals. We should expect that the function point of view gives a more accurate understanding provided that the structure of $G$ is more accurate. On the other hand, the geometric point of view provides us with a means to enhance our knowledge of the graph structure. In summary, we want to explore geometric information contained in graph signals as a central theme. 

By \cref{eg:cah}, we see describing non-smooth graph signals as noise can cause information loss. The seemingly noisy signal in the example is in the perpendicular direction to the planar embedding of the graph. This prompts us to introduce the notion of perpendicular graph signals. 

\begin{Definition} \label{def:lsb}
Let $S$ be a subset of graph signals. Then the space of \emph{perpendicular graph signals w.r.t.\ $S$} is the orthogonal complement $S^{\perp}$ of $S$. 
\end{Definition}

Though we have a seemingly naive notion here, the real point is that the set $S$ contains those signals coherent with the graph structure, such that it makes sense to analyze them with current GSP techniques. Such a signal is considered as ``smooth'', which will be made explicit in \cref{sec:smo}. On the other hand, a signal $f$ in $S^{\perp}$ contains geometric information ``perpendicular'' to that offered by $G$ (c.f.\ \cref{eg:cah}). Hence, we may consider using $f$ to enhance our understanding of $G$, instead of processing it with filters built from $G$. According to \cref{def:lsb}, smooth signals and perpendicular signals are the two sides of the same coin. Our strategy is to focus on the former explicitly in \cref{sec:smo}.

We end this section by providing further intuitions and heuristics. We describe some major challenges to ``compare the geometry of graph and signal'', which motivate subsequent sections. 

To proceed, we introduce a notion here. Suppose $f=(f_i)_{1\leq i\leq n},g=(g_i)_{1\leq i\leq n} \in \mathbb{R}^n$. For convenience, the notation is only used till the end of this subsection. We say that $f$ \emph{interlaces} $g$ if there is a constant $c$ such that $h=f+c$ satisfies $g_i\leq h_i\leq g_{i+1}$ for $1\leq i\leq n-1$, $h_n\geq g_n$ and $h_1g_1+h_ng_n\geq 0$. Notice that the last condition loosely controls the absolute values of $g_1$ and $h_n$.

\begin{Lemma}\label{lem:fns}
For $n\geq 3$, suppose $f, g\in \mathbb{R}^n$ are nonzero vectors and $g$ is orthogonal to constant vectors. If a re-arrangement of indices makes $f$ interlace $g$, then $f$ and $g$ are not orthogonal to each other.  
\end{Lemma}

Let us try to interpret ``$f$ interlaces $g$'' geometrically. According to the definition, the increments in the signal values $f_{i+1}-f_i$ are controlled by those of $g$. In particular, if the increments are small for $g$, so are those for $f$. On the other hand, if $f$ does not interlace $g$, then we can observe large increments for $f$ can occur when we have a small increment in $g$. In the GSP setting, one argues heuristically that a smooth signal is the one that has a small increment across many direct edges. Therefore, if we have a set $S$ exhausts smooth signals, then any $f$ perpendicular to $S$ does not interlace any of the signals in $S$. Based on our discussions, such $f$ tends to violate having small increments along a direct edge for many different edges.

However, to make our discussion rigorous, we need to specify what we mean by a signal being smooth, aside from the hand-waving description as above. The main challenge is that signals and graphs are different mathematical objects. Therefore, they are not directly comparable. Our key task is to set-up a common platform so that we can discuss the geometric contents of both graphs and signals and make comparisons. 

\section{Graph-signal coupling} \label{sec:cou}

As we have pointed out in the previous section, a graph signal can contain geometric information not captured by the graph. In this section, we shall discuss a framework, called \emph{graph-signal coupling}, to extract geometric information from graphs and signals combined. 

\subsection{An axiomatic approach}

We want to first propose an axiomatic approach to avoid restricting to a single construction, while we still give an explicit construction in the next subsection. The axiomatic approach proceeds by stating a few desired properties any construction needs to satisfy. We start by motivating such properties.

Recall that we want to find a ``common platform'' to compare graphs and signals. More specifically, let $\mathcal{S}$ be a set of objects having geometric interpretation. For example, $\mathcal{S}$ can be the set of graphs or the set of finite (pseudo) metric spaces (recall in a pseudo metric space, $d(x,y)=0$ does not imply $x=y$). Suppose $F = \{f_1,\ldots, f_m\}$ contains a finite set of signals on $G$. We want to produce an object $G_F \in \mathcal{S}$.

First of all, we \emph{need a notion of equivalence} between objects in $\mathcal{S}$, and the reason is as follows. Suppose $G_1$ and $G_2$ are isomorphic graphs by a permutation $\sigma$ of the nodes. For $F_1$ on $G_1$, permuting (with $\sigma$) the indices of each signal in $F_1$ results a set of signals $F_2= \sigma(F_1)$. The objects ${G_1}_{F_1}$ may not be the same as ${G_2}_{F_2}$, but they should be equivalent.

The coupling should be extendable to pairs $G_{F_1} \in \mathcal{S}$ and $F_2$, where both $F_1$ and $F_2$ are finite sets of signals on $G$. More precisely, we want to have an object $(G_{F_1})_{F_2} \in \mathcal{S}$. This is because regarded as observations, the signals $F_1$ and $F_2$ may not be obtained simultaneously. Moreover, we also want that $G_F \in \mathcal{S}$ allows us to recover both the graph and signals to a certain extent. 

Keeping these requirements in mind, we now formalize the idea of graph-signal coupling.

\begin{Definition} \label{defn:lgb}
Let $\mathcal{G}$ be the set of graphs. A \emph{graph-signal coupling} consists of the following data: a set of objects $\mathcal{S}$ and an equivalence relation ``$\sim$'' on $\mathcal{S}$. For each graph $G$ and a finite set of signals $F$, there is a $G_F \in \mathcal{S}$ (if $F=\{f\}$, we write $G_f$ for $G_F$ for convenience) such that the following holds:
\begin{enumerate}[1)]
    \item Composability: For each finite set of signals $F'$, there is $(G_F)_{F'} \in \mathcal{S}$. 
    \item\label{it:cna} Commutativity and associativity: $(G_F)_{F'}\sim G_{F\cup F'}$.
    \item Recoverability of $G$: There is \emph{base map} $b: \mathcal{S} \to \mathcal{G}$ such that $b(G_F) = G$.
    \item Recoverability of signal: $G_f = G_{f'}$ implies $|f(v_i)-f(v_j)| = |f'(v_i)-f'(v_j)|, 1\leq i,j\leq n$. Moreover, if $f = af'+c$ for scalar $a\neq 0$ and constant signal $c$, then $G_f\sim G_{f'}$.  
\end{enumerate}
\end{Definition}

Before presenting an explicit construction in \cref{sec:exp}, we analyze the obvious choice of $\mathcal{S}$ being the collection of graphs. This means for a graph $G$ and a finite set of signals $F$, we need to produce a new graph $G_F$. In the collection of graphs, the most reasonable notion of equivalence is graph isomorphism. Two isomorphic graphs are essentially the same up to a re-ordering of vertices. In addition, for any $G$ and a constant signal $c$, it is reasonable to require $G_c$ isomorphic to $G$. Consider any signal $f$. Condition~\ref{it:cna} forces $(G_f)_f$ isomorphic to $G_f$. The latter in turn is isomorphic to $(G_f)_c$ for any constant signal. This means that on the graph $G_f$, we are not able to differentiate $f$ from any constant signal, i.e, the recoverability of signals is violated.

If we examine the above argument, we notice that the cause of the problem is that equivalence on our current choice of $\mathcal{S}$ is too restrictive. As a remedy, we shall consider that $\mathcal{S}$ contains a parametrized family of graphs in \cref{sec:exp}.

\subsection{An explicit construction} \label{sec:exp}

In this subsection, we give an explicit construction that verifies the properties listed in \cref{defn:lgb}. The construction is inspired by the basic construction in linear algebra: taking the sum of perpendicular vectors. 

In the construction, an object in $\mathcal{S}$ is a parametrized family of graphs. More precisely, it is a map $\gamma: M \to \mathcal{G}$, where $M$ is a parameter space such as a topological space or a manifold, and $\mathcal{G}$ is the collection of finite graphs. For convenience, we use $\gamma$ to denote such an object.

\begin{figure}[!htb] 
	\centering
	\includegraphics[scale=0.5]{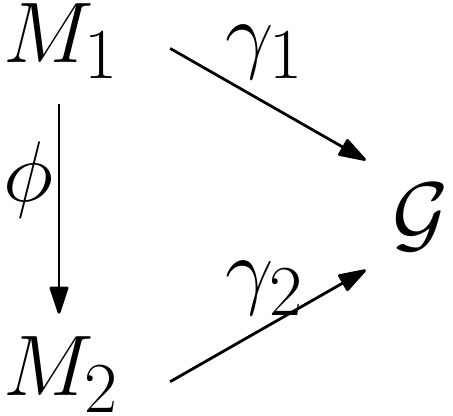}
	\caption{Schematic diagram of a morphism.}\label{fig:20}
\end{figure}

Given $\gamma_i: M_i \to \mathcal{G}, i=1,2$, a \emph{morphism} from $\gamma_1$ to $\gamma_2$ is a map $\phi: M_1 \to M_2$ such that $\gamma_1 = \gamma_2\circ \phi$ (illustrated in \figref{fig:20}). If $M_i$ are topological spaces, we usually require $\phi$ to be continuous; and if $M_i$ are differentiable manifolds, we want $\phi$ to be differentiable. 
To understand this notion, consider $g\in Im(\gamma_1)$, i.e., $g = \gamma_1(x)$ for $x\in M_1$. Then $y = \phi(x)$ satisfies $\gamma_2(y) = g$. Intuitively, this indicates that $\phi$ is analogous to a ``surjection'' from $\gamma_2$ to $\gamma_1$, though we notice a reverse of domain and codomain. Now we proceed to define the equivalence relation. 

\begin{Definition}
Given $\gamma_1,\gamma_2$, we write $\gamma_1 \leq \gamma_2$ if there is a morphism $\phi$ from $\gamma_1$ to $\gamma_2$. We say $\gamma_1$ and $\gamma_2$ are equivalent to each other, denoted by $\gamma_1 \sim \gamma_2$ if $\gamma_1\leq \gamma_2$ and $\gamma_2\leq \gamma_1$.
\end{Definition}

The equivalence is based on the analogy that two finite sets are equivalent to each other if either surjects onto the other. 

We can now define $G_F$ for given graph $G$ and a finite set of signals $F = \{f_1,\ldots, f_k\}$. Let $M_F = \mathbb{R}_{\geq 0} ^{k+1}$. We write $x = (x_0,x_1,\ldots,x_k)$ for a typical element in $M_F$. 

Then $G_F \in \mathcal{S}$ is a map $G_F: M_F \to \mathcal{G}$ defined by\footnote{The construction is inspired by the idea of \cite{Carl09} Definition 2.5.}
\begin{enumerate}[1)]
    \item For two nodes $u,v$, compute
    \begin{align*}
        \Delta(u,v) = \Big(\sum_{1\leq i\leq k}x_i(f_i(u)-f_i(v))^2 + d_G(u,v)^2\Big)^{1/2},
    \end{align*}
    where $d_G$ is the distance on $G$.
    \item In $G_F(x)$, a pair of nodes $u,v$ is connected by an edge if $\Delta(u,v)\leq x_0$. 
\end{enumerate}

Intuitively, $x_i, 1\leq i\leq k$ scales the difference measured by the signal $f_i$. The parameter $x_0$ is a threshold to determine the connections in $G_F(x)$. Of course, we need to check the following.

\begin{figure}[!htb] 
	\centering
	\includegraphics[scale=0.75]{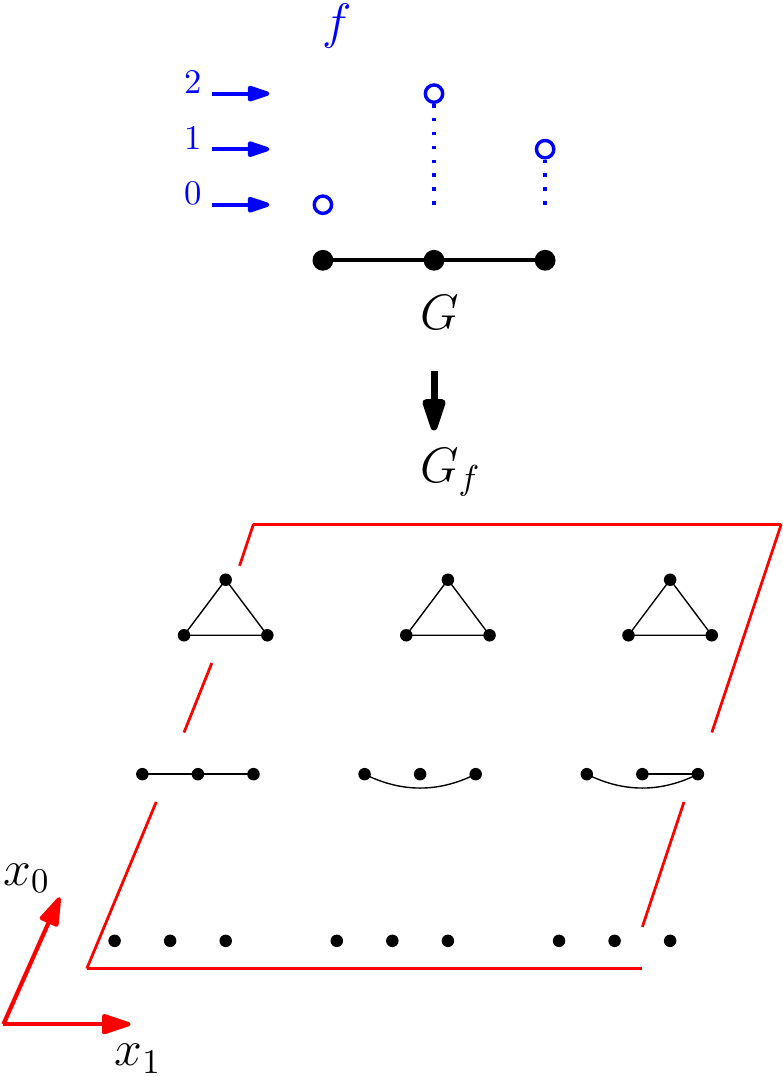}
	\caption{In this example, $G$ is the path graph on $3$ nodes, and the signal $f$ is $(0,2,1)^T$. As shown below, $G_f$ is the parameter family of graphs. There are two parameters $(x_0,x_1)$. The parameter $x_1$ scales signal differences and $x_0$ serves as a threshold to determine edge connections.}\label{fig:19}
\end{figure}	

\begin{Theorem} \label{prop:agw}
Associating $G,F$ with $G_F: M_F \to \mathcal{G}$ is a graph-signal coupling.
\end{Theorem}

\subsection{The abstract picture} 

In this section, we give a more abstract description of the picture, to explain the theoretical underpinning of our approach.

Let $\mathcal{G}_n$ be the collection of unweighted, undirected finite graphs on $n$ vertices. We start with an object $G$ in this collection. On the other hand, we want to study signals on such a graph $G$. The collection of such signals can be identified with $\mathbb{R}^n$, denoted by $\mathcal{F}_n$. As we propose in this paper, we want to investigate geometric information contained in a signal $f \in \mathbb{R}^n$. While a direct comparison of $f$ and $G$ is obscure as they belong to different collections of objects, we want a common, new collection of objects that enlarges both $\mathcal{G}_n$ and $\mathcal{F}_n$. 

\begin{Proposition} \label{prop:imt}
Introduce 
\begin{itemize}
\item $\mathcal{M}_n$: the collection of metric spaces of size $n$; 
\item $\mathcal{MG}_n$: the collection of map $\phi: M \to \mathcal{G}_n$, i.e., as a parametrized family of graphs, by a topological space $M$ such that $Im(\phi)$ is finite; and
\item $\mathcal{SG}_n$: the collection of a finite sequence of undirected, unweighted graphs of size $n$.
\end{itemize}
Then we have the diagram of maps as shown in \figref{fig:18}
\begin{figure}[!htb] 
	\centering
	\includegraphics[scale=0.6]{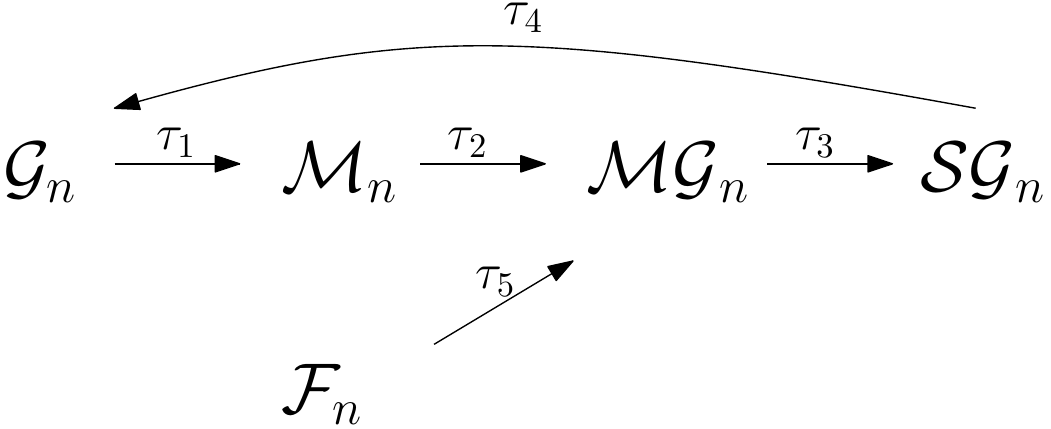}
	\caption{The schematic illustration of \cref{prop:imt}.}\label{fig:18}
\end{figure}	
such that $\tau_4\circ \tau_3 \circ \tau_2 \circ \tau_1 = Id_{\mathcal{G}_n}$, $\tau_2\circ\tau_1(G) = G_c$ for any constant signal $c$. Moreover, if $\tau_5(f)=\tau_5(g)$, then $f = g + c$ or $f=-g+c$ for some constant signal $c$. 
\end{Proposition}

As a consequence, we have placed both graphs and signals in the common collections $\mathcal{MG}_n$ and $\mathcal{SG}_n$, where we can compare them  directly. 

\section{Smooth graph signal} \label{sec:smo}

We have introduced perpendicular signals earlier, which contain additional geometric information. It is supposed to come together with the notion of smooth signals, which we formally define in this section. The main idea is that we make use of the constructions of \cref{sec:cou} to place both graph and a signal in a common collection of objects and make comparisons based on an appropriate measure defined on the collection. The first goal is to introduce such a measure.

Let $*$ be the trivial graph on a single node. We also view it as a degenerate graph on $n$ nodes by identifying all of them with the single node, i.e., the distance between any pair of nodes is $0$. On the other hand, on the graph $G$, let $c$ be the unit constant signal.

Recall that according to \cref{sec:cou}, we have defined $*_f$ and $G_c$ as objects in some collection $\mathcal{S}$. By \cref{prop:imt}, $\mathcal{S}$ can be either the collection of a parameterized family of graphs or the collection of a finite sequence of graphs.

\begin{Definition} \label{def:att}
Assume that there is $d_{\mathcal{S}}: \mathcal{S}\times \mathcal{S} \to \mathbb{R}_{\geq 0}$. Then for $\epsilon\geq 0$, a graph signal $f$ on $G$ is called $\epsilon$-smooth (w.r.t,\ $d_{\mathcal{S}}$) if $d_\mathcal{S}(*_f, G_c)\leq \epsilon$.
\end{Definition}

As we pointed out earlier, we want $d_{\mathcal{S}}$ to play the role of a metric to measure how different two elements of $\mathcal{S}$ are. However, the $d_{\mathcal{S}}$ we are going to define does not satisfy all the properties of a metric such as being symmetric.

We now describe an explicit construction of $d_{\mathcal{S}}$ based on our choice of $\mathcal{S}$ and construction of $G_F$. We only consider elements of $\mathcal{S}$ taking the form of $G_F$ (cf.\ \cref{sec:cou}). Recall that for two unweighted graphs $G_1$ and $G_2$ on the same amount of $n$ vertices, their \emph{Hamming distance} $d_H(G_1,G_2)$ counts the number of edges contained exclusively in either $G_1$ or $G_2$. More generally, if $\Gamma$ is a set of unweighted graphs on $n$ vertices, then \begin{align*} d_H(G_1, \Gamma) = \min_{G_2\in \Gamma}d_H(G_1,G_2). \end{align*}

For $G$ of size $n$ and a finite set of graph signals $F$, we have constructed $G_F: M_F \to \mathcal{G}$ such that $M_F$ is a subset of a Euclidean space. The image $Im(G_F)$ is finite. An extreme case in $Im(G_F)$ is the the complete graph $K_n$, and write $Im(G_F)^{\circ}$ for $Im(G_F)\backslash \{K_n\}$. Let the inverse image of $Im(G_F)^{\circ}$ has \emph{Lebesgue measure} $|G_F^{-1}(Im(G_F)^{\circ})|$. 

\begin{figure}[!htb] 
	\centering
	\includegraphics[scale=0.8]{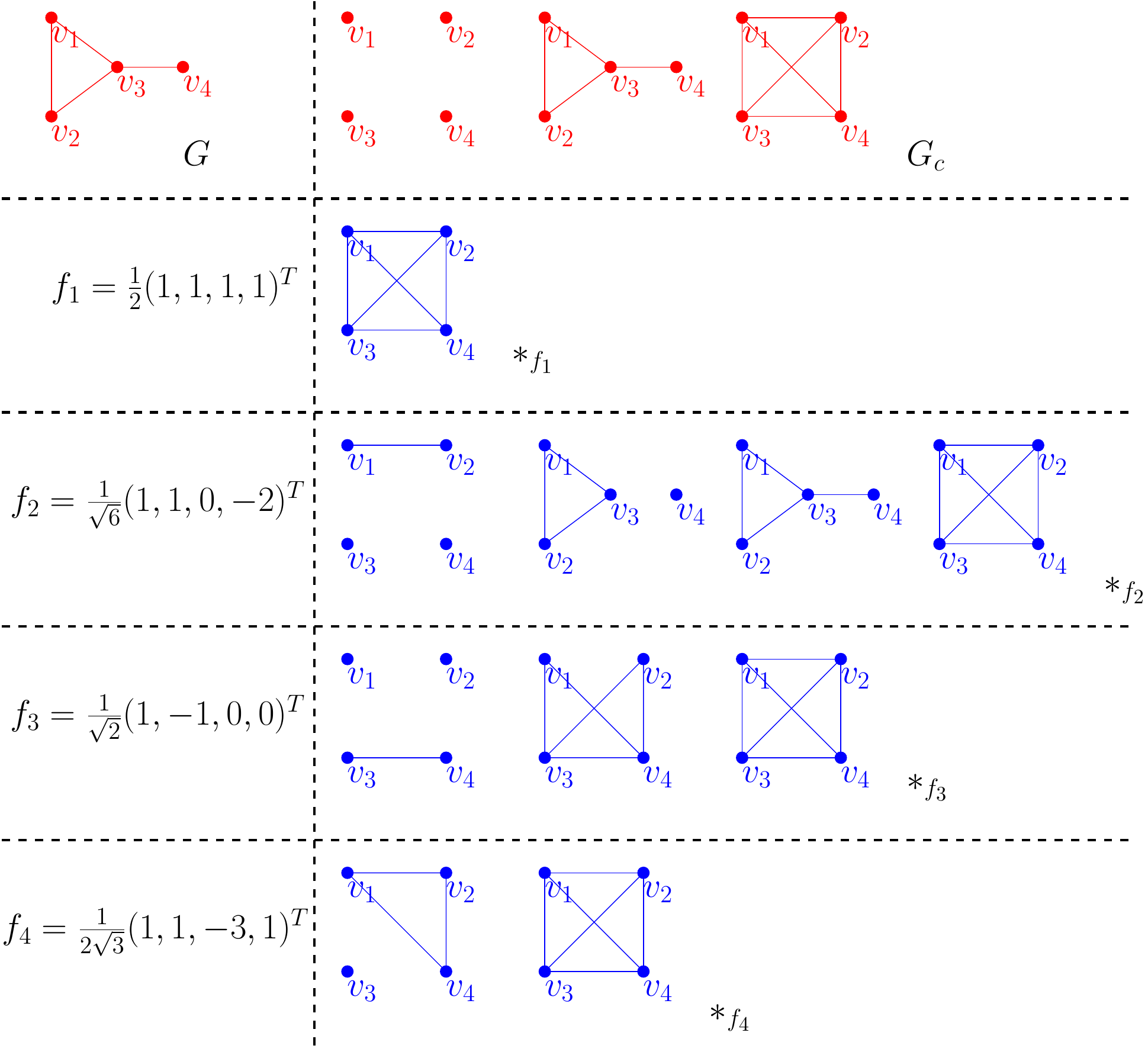}
	\caption{In this example, $G$ is a graph on $4$ nodes. In the top row, we show the image of $G_c$ consisting of a sequence of $3$ distinct graphs. On the other hand, we consider an eigenbasis $\{f_1,f_2,f_3,f_4\}$ of the Laplacian of $G$. In the figure, we show the image of $*_{f_i}, 1\leq i\leq 4$ in subsequent rows. Using the formula for $d_{\mathcal{S}}(\cdot,\cdot)$, we find that $d_{\mathcal{S}}(*_{f_i},G_c), 1\leq i\leq 4$ are $0, 2/3, 1, 3$ respectively.}\label{fig:17}
\end{figure}

Suppose $G_1$ and $G_2$ are graphs on the same set of ordered vertices $V$ and finite sets of graph signals $F_1$ and $F_2$. For convenience, denote $\gamma_1 = {G_1}_{F_1}$ and $\gamma_2= {G_2}_{F_2}$. If $Im(\gamma_1)^{\circ}$ is non-empty define 
\begin{align} \label{eq:dms}
    d_{\mathcal{S}}(\gamma_1,\gamma_2) = \sum_{G\in Im(\gamma_1)^{\circ}} \frac{|\gamma_1^{-1}(G)|}{|\gamma_1^{-1}(Im(\gamma_1)^{\circ})|}d_H(G,Im(\gamma_2)),
\end{align}
where $|\cdot|$ denotes the Lebesgue measure of the set. If $Im(\gamma_1)^{\circ}=\emptyset$, then $d_{\mathcal{S}}(\gamma_1,\gamma_2)=0$. An illustration is shown in \figref{fig:17}.

We now study some basic properties of $\epsilon$-smooth signals.

\begin{Lemma} \label{lem:ifi}
\begin{enumerate}[1)]
    \item If $f$ is $\epsilon$-smooth, then so are $rf$ and $f+c$ for any $r\in \mathbb{R}\backslash\{0\}$ and $c$ a constant signal.
    \item If $G$ is connected, then the set of $0$-smooth signals contains only constant signals if and only if $G$ is not a path.
    \item \label{it:tmm} The map $\mathbb{R}^n \to \mathbb{R}, f \mapsto d_\mathcal{S}(*_f, G_c)$ is continuous on the subset of non-constant signals.
\end{enumerate}
\end{Lemma}

We now discuss other means to estimate the parameter $\epsilon$ without directly using the definition.

\begin{Example} \label{eg:wrf}
In this example, we re-visit \figref{fig:17}. In particular, we want to compute $d_{\mathcal{S}}(\gamma_1,\gamma_2)$ for $\gamma_1=*_{\sqrt{6}f_2} = *_{f_2}$ and $\gamma_2 = G_c$, without directly using the definition. The idea here is that instead of considering each graph of $Im(*_{f_2})$, we consider edges in these graphs. More specifically, we order pairs of distinct vertices $\{u,v\}$ increasingly according to $\Delta(\{u,v\}) = |f_2(u)-f_2(v)|$. We obtain the sequence: $Q = (Q_1,\ldots, Q_6): \{v_1,v_2\}, \{v_1,v_3\}, \{v_2,v_3\}, \{v_3,v_4\}, \{v_1,v_4\}, \{v_2,v_4\}$ with the corresponding $\Delta$ values $0,1,1,2,3,3$. We partition $Q$ into $3$ parts: $I_0=(Q_1), I_1=(Q_2,Q_3,Q_4)$ and $I_2=(Q_5,Q_6)$.

Notice that in \figref{fig:17}, we see that $Im(G_c)$ contains $3$ graphs, and we call them $G^{(0)}$, $G^{(1)}$ and $G^{(2)}$ respectively. 

Now we process each $Q_l,l=1,\ldots,6$ to obtain a number $\epsilon_l$. We show how this is done for typical examples $Q_1$ and $Q_4$. For $Q_1=\{v_1,v_2\}$, we notice that the pair first appears as an edge in $G^{(1)}$. We then identify the first pair of $I_1$ is $Q_2=\{v_1,v_3\}$. Then we compute $ \epsilon_1=|\Delta(Q_1)-\Delta(Q_2)|/\Delta(Q_6)=1/3$. Similarly for $Q_4 = \{v_3,v_4\}$, we notice that it first appears as an edge in $G^{(1)}$ as well, and determine $Q_2$ being the first pair of $I_1$. We then find $\epsilon_4 = |\Delta(Q_4)-\Delta(Q_2)|/\Delta(Q_6)=1/3$. The same computation determines that $\epsilon_2=\epsilon_3=\epsilon_5=\epsilon_6=0$. Summing all of them we obtain $\sum_{1\leq l\leq 6} \epsilon_6=2/3$, which is exactly the smoothness of $f_2$. 

We shall next rigorously describe and then demystify the procedure.
\end{Example}

Before formalizing the procedure in \cref{eg:wrf}, we make the following technical assumption on $f$: for two different pairs (as sets) of nodes $\{u_1, v_1\}$ and $\{u_2, v_2\}$, we have $|f(u_1)-f(v_1)| \neq |f(u_2)-f(v_2)|$.

For the graph $G$, let $D_G$ be its diameter. For each $0\leq k\leq D_G$, we define $G^{(k)}$ be to the graph on $V$ and $u, v$ is connected by an edge if and only if $d_G(u,v)\leq k$. This sequence of graphs $G^{(k)}, 0\leq k\leq D_G$ is nothing but the image of $G_c$. 

Suppose we order the distinct pairs of vertices $\{u,v\}$ of $V$ increasingly according to $|f(u)-f(v)|$. In this way, we obtain an ordered sequence of pairs of vertices $Q=(Q_l)_{1\leq l\leq n(n-1)/2}$, with $Q_l=\{u_l,v_l\}$. A \emph{$D_G+1$-partition} $\mathcal{P}$ of $I =\{1,\ldots, n(n-1)/2\}$ is a decomposition of $I = \cup_{0\leq k\leq D_G}I_k$ into $D_G+1$ disjoint subsequences of consecutive numbers $I_k$ in $I$. We assume that the numbers in $I_k$ are smaller than those in $I_{k+1}$, and we allow $I_k$ to be empty. We compute $\epsilon_\mathcal{P}$ as follows. 

\begin{algorithm}[!htb]
\caption{}\label{alg:rau}
\hspace*{\algorithmicindent} \textbf{Input}: $G, f, Q, \mathcal{P}$
\hspace*{\algorithmicindent} \textbf{Output}: $\epsilon_{\mathcal{P}}$ of $f$
\begin{itemize}
\item For each $1\leq l \leq n(n-1)/2$, let $k$ be the index such that $(u_l,v_l)$ is an edge of $G^{(k)}$ for the first time, in the sequence $G^{(0)}, \ldots, G^{(D_G)}$.
\item Let $l'$ be first element of $I_k$ and $Q_{l'}=\{u_{l'},v_{l'}\}$.
\item Set
\begin{align*}
\epsilon_l = \frac{\big| |f(u_{l'})-f(v_{l'})|-|f(u_l)-f(v_l)|\big|}{|f(u_{(n(n-1)/2)})-f(v_{n(n-1)/2})|}.
\end{align*}
\item Summing over $1\leq l\leq n(n-1)/2$, \begin{align*}\epsilon_\mathcal{P}=\sum_{1\leq l\leq n(n-1)/2} \epsilon_l.\end{align*}
\end{itemize}
\end{algorithm}

The observation made in \cref{eg:wrf} is demystified by the following result.

\begin{Proposition} \label{prop:leb}
Let $\epsilon$ be the smallest number such that $f$ is $\epsilon$-smooth. Then $\epsilon = \min_{\mathcal{P}}\epsilon_{\mathcal{P}}$, where the minimum is taken over all partitions $\mathcal{P}$ of $I=\{1,\ldots, n(n-1)/2\}$ into $D_G+1$ subsequences.
\end{Proposition}

Based on the proof of \cref{prop:leb}, we can describe the role played by the partition $\mathcal{P} = \{I_0,\ldots, I_{D_G}\}$. To compute $\epsilon_{\mathcal{P}}$, we form the sequence of graphs $G_{(l)}$ by including $Q_1,\ldots, Q_l$ as edges. In the expression $d_H(\cdot,\cdot)$, we put $G_{(l)}$ with $G^{(k)}$ as the arguments for $l\in I_k$. By the proposition, any choice of $\mathcal{P}$ allows us to obtain an upper bound of $\epsilon$. Moreover, to find the optimal partition, we only need to find the starting and ending indices of each $I_k, 0\leq k\leq D_G$. This leads to \cref{alg:gfp} that is based on the binary search of such indices.

\begin{algorithm}[!htb] 
\caption{}\label{alg:gfp}
\hspace*{\algorithmicindent} \textbf{Input}: $G, f$
\hspace*{\algorithmicindent} \textbf{Output}: $\mathcal{P}$ of $f$
\begin{itemize}
\item Construct $G^{(k)}$ for $0\leq j \leq D_G$.
\item Form the sequence of pairs of nodes $Q = (Q_l)_{1\leq l\leq n(n-1)/2}$ with $Q_l=\{u_l,v_l\}$ such that $|f(u_l)-f(v_l)|<|f(u_{l+1})-f(v_{l+1})|$.
\item For any $Q_l$, recall $G_{(l)}$ is the graph with edges $Q_1,\ldots, Q_l$.
\item For each $0\leq k\leq D_G$, apply binary search to $Q$ to find the starting and ending indices of $I_k$, with the following rule:  Then $l \in I_k$ if $k$ is the largest element in $\argmin_{0\leq j\leq D_G}d_H(G_{(l)},G^{(j)})$. Any index $l$ such that $Q_l$ is considered in each iteration is used as a reference index in subsequent iterations.
\end{itemize}
\end{algorithm}

To end this section, we slightly generalize \cref{def:att}. What is missing from the current notion of $\epsilon$-smoothness is that it is not preserved under vector addition. Hence, in general, they do not form a vector space. This is unfavorable in signal processing. On the other hand, \cref{lem:ifi}~\ref{it:tmm} shows that if $\epsilon>0$, in general $\epsilon$-smooth vectors span $\mathbb{R}^n$. To come up with useful vector spaces, we propose the following.

\begin{Definition}
Suppose $F = \{f_1, \ldots, f_n\}$ is an orthonormal basis of $\mathbb{R}^n$. Then let $F_{\epsilon} = \{f \in F \mid f: \epsilon\text{-smooth}\}$. An arbitrary signal is $\epsilon$-smooth w.r.t.\ $F$ if $f$ is in the span of $F_{\epsilon}$.

If $L$ is a matrix admitting an eigenbasis, then $f$ is $\epsilon$-smooth w.r.t.\ $L$ if it is $\epsilon$-smooth w.r.t.\ an eigenbasis of $L$.

More generally, $f$ is $\epsilon$-smooth if it is in the span of a set of pairwise orthogonal $\epsilon$-smooth vectors.
\end{Definition}

\section{Simulations} \label{sec:sim}

In this section, we provide simulation results. The main focus is to demonstrate that well-established GSP tools can be modified with our new framework to give new insights and experimental observations. 

\subsection{Band-pass filters: old wine in new bottles}

In this subsection, we study band-pass filters. We first recall briefly what they are in GSP. Let $L$ be a fixed GSO such as the Laplacian of $G$. It admits an orthonormal eigenbasis $\mathcal{E} = \{e_1,\ldots, e_n\}$ with the corresponding eigenvalues $0=\lambda_1 \leq \lambda_2 \leq \ldots \leq \lambda_n$. For a subset $Y$ of $[n] = \{1,\ldots, n\}$, the \emph{band-pass filter $B_Y$} on any graph signal $f$ is defined as 
\begin{align*}
B_Y(f) = \sum_{i\in Y} \langle e_i,f\rangle e_i,  
\end{align*}
where $\langle \cdot, \cdot \rangle$ is the standard inner-product. The band-pass filter retains only the components, indexed by $Y$, of the eigen-decomposition of $f$. A slightly more general version is that we take a pair of numbers $\balpha = (\alpha_0, \alpha_1)$, and construct $B_{Y,\balpha} = \alpha_0B_Y + \alpha_1I_n$, with $I_n$ the identity map. For example, $B_Y = B_{Y,\balpha}$ for $\balpha=(1,0)$. The coefficients $\balpha$ provide additional flexibility if we do not want to completely disregard contributions from $e_i$ for $i\notin Y$.

In many tasks, the index set is chosen as $X_m = \{1,\ldots, m\}$ for some $m<n$. The resulting filter is called a low-pass filter. It leverages the intuition that structured signals are smooth in the sense that it contains mainly ``low frequency'' components. Here, ``low'' refers to small eigenvalues. In our paper, we provide a different interpretation of smoothness as formally defined in \cref{sec:smo}. This allows us to choose an index set according to the smoothness therein. We describe how this simple procedure is done as follows.

\begin{algorithm}[!htb]
\caption{}\label{alg:emy}
\hspace*{\algorithmicindent} \textbf{Input}: $\mathcal{E}, G, m$
\hspace*{\algorithmicindent} \textbf{Output}: $Y_m$ of size $m$
\begin{itemize}
\item For each $e_i \in \mathcal{E}$, find $\epsilon_i = d_H(*_{e_i},G_c)$ for any constant signal $c$.
\item Determine a permutation $\sigma$ of $1,\ldots, n$ according to increasing order of $\epsilon_i$, i.e., $\epsilon_{\sigma(i)} \leq \epsilon_{\sigma(i+1)}$.
\item $Y_m = \{\sigma(1), \ldots, \sigma(m)\}$.
\end{itemize}
\end{algorithm}

The index set $Y_m$ constructed in \cref{alg:emy} shall be used as a substitute of $X_m=\{1,\ldots, m\}$ in classical GSP. In general, the resulting filter $B_Y$ is not a low-pass filter in the classical sense. We shall demonstrate with simulations.  

\begin{figure}[!htb] 
	\centering
	\includegraphics[trim={0 6.5cm 0 6.5cm}, clip, scale=0.45]{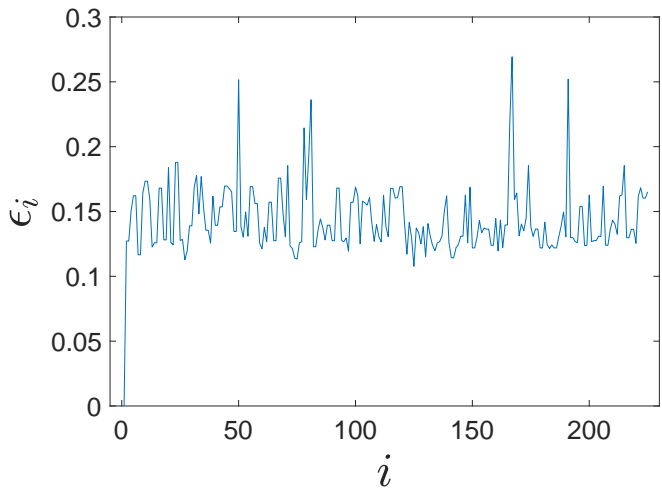}
	\caption{The plot of $\epsilon_i$ against $i$, where the indices on the horizontal axis are ordered according to the graph frequencies. }\label{fig:lattice_plot}
\end{figure}

We consider the MNIST dataset.\footnote{http://yann.lecun.com/exdb/mnist/} We use a $2$D-lattice $G=(V,E)$ to model the graph for each image, and $L$ is the Laplacian of $G$. As described above, it has an orthonormal eigenbasis $\mathcal{E} = \{e_1,\ldots, e_n\}$. We first compute $\epsilon_i$ as in \cref{alg:emy}, to investigate the relation between the notions of smoothness introduced in the paper and implicitly suggested by classical GSP. In \figref{fig:lattice_plot}, we show the plot of $\epsilon_i$ (normalized by the size of $G$) against $i$. The indices on the horizontal axis are ordered according to the sizes of the eigenvalues of $L$. Therefore, from the plot, we see that an index $i$ with small GSP frequency does not necessarily have small $\epsilon_i$. As a consequence, $Y_m$ can be very different from the set $X_m = \{1,\ldots, m\}$, which is used to construct a low-pass filter in classical GSP. A low-pass filter in terms of size of $\epsilon_i$ is a classic band-pass filter, but not necessarily a classical low-pass filter. 

\begin{figure}[!htb] 
	\centering
	\includegraphics[trim={1cm 6cm 0 4cm}, clip, scale=0.75]{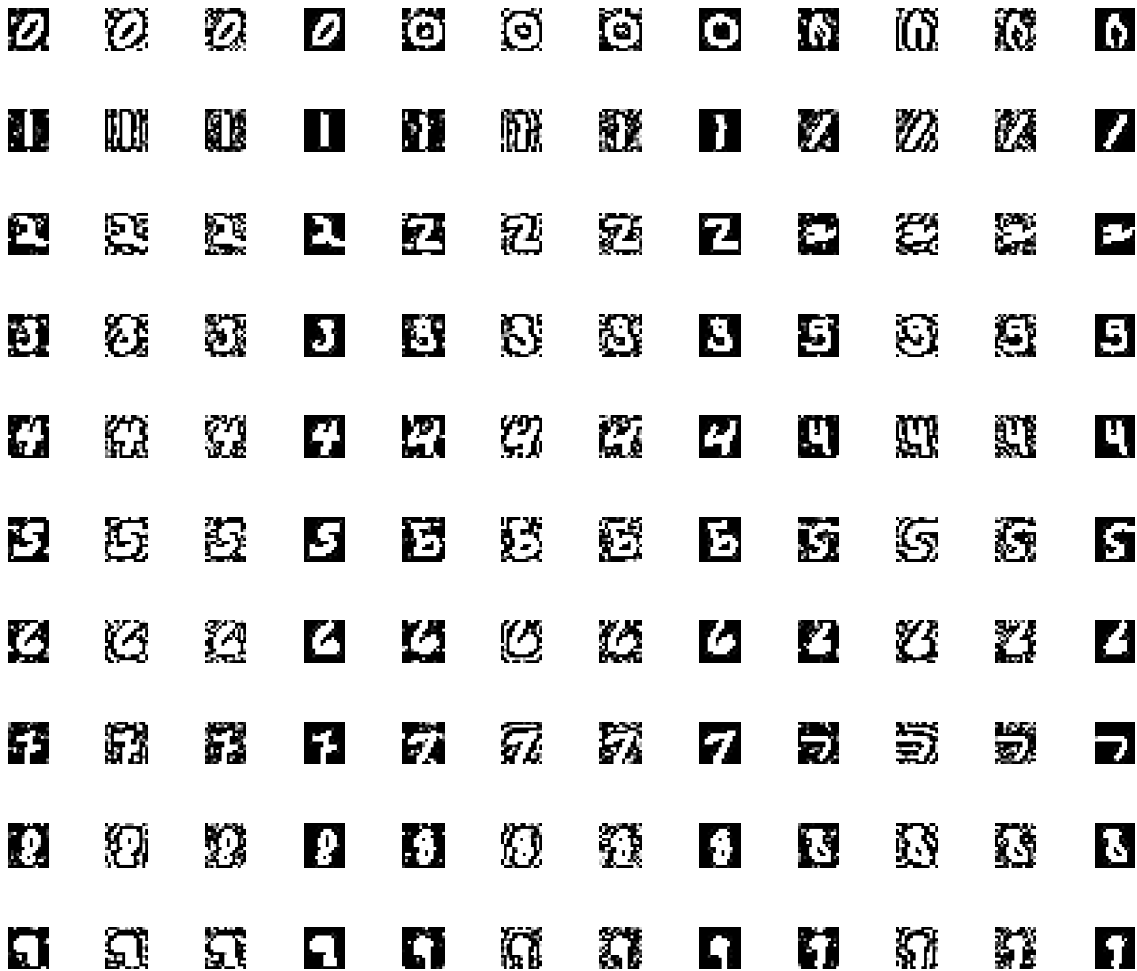}
	\caption{We show sample images recovered from noisy ones using $B_{X_m,\balpha}$ and $B_{Y_m,\balpha'}$. Each row corresponds to a digit from $0$ to $9$. Every four columns consist of an example. In each example, the last image of the group is the original image and the third image is the noisy image. The first image is the one obtained by applying $B_{Y_m,\balpha'}$, while the second image is obtained by applying $B_{X_m,\balpha}$.  }\label{fig:mnist}
\end{figure}

We investigate the difference between the band-pass filters with $X_m = \{1,\ldots, m\}$ and $Y_m = \{\sigma(1),\ldots, \sigma(m)\}$ on processing noisy images, with $m \approx 0.2n$. To do so, we add independent Gaussian noise to each pixel of the samples in the MNIST dataset. We apply appropriate band-pass filters, as denoising functions, to the noisy images to recover the original images. To be flexible, the filters are $B_{X_m,\balpha}$ and $B_{Y_m,\balpha'}$ as described at the beginning of this subsection. Both coefficient sets $\balpha$ and $\balpha'$, as hyperparameters, are tuned based on a small number of samples. In \figref{fig:mnist}, we show the results. We see that with $B_{Y_m,\balpha'}$, the recovered images look more like the original ones (especially the complement of the digit figures) as compared with those recovered with $B_{X_m,\balpha}$, which is the classical low-pass filter.

\subsection{Principle components of graph signals}

In this subsection, we investigate the construction of \cref{sec:cou} by studying the principle components of graph signals. The dataset we use here is from a weather station network in the United States with $n=197$ nodes.\footnote{\url{http://www.ncdc.noaa.gov/data-access/}} The signals are daily temperatures recorded over the year 2013.

We pre-process each $f$ in our dataset as follows. Let $c$ be any constant signal, we place $f$ by $f-\langle f,c\rangle c$, i.e., the constant component from $f$ is removed as it does not give any useful geometric information of the graph. The pre-processing step results in a perpendicular signal to the $0$-smooth signals according to \cref{def:lsb}.

The geographic locations of the stations are available, and based on such information, a $k$-NN graph $G=(V,E)$ is constructed. The size of the graph is $|V|=194$, and the average degree of $G$ is $\approx 5$ and $|E|=495$. Among all the signals, we randomly choose $3$ samples from each month to form a sample set of signals $F$ containing $36$ signals, approximately $10\%$ of all the available signals. 

By the construction of \cref{sec:cou}, we obtain a parametrized family of graphs $G_F: M_F \to \mathcal{G}$, where $M_F = \mathbb{R}_{\geq 0}^{37}$ and $\mathcal{G}$ is the collection of graphs on $194$ vertices. The entire image $Im(G_F)$ of $G_F$ is too large for us to investigate. For our purpose, we consider a subset $C = \{G_0,\ldots, G_{20}\}$ of $Im(G_F)$ consisting of $21$ graphs. Each $G_i, 0\leq i\leq 20$ is of the form $G_F(y_i,x_i,\ldots, x_i)$ such that:  
\begin{itemize}
    \item $x_0=0$ (hence $G_0 = G$).
    \item $x_{20}>\!\!> 1$ and $x_1 <\ldots< x_{19}$ are chosen to equally divide the interval $(x_0,x_{20})$.   
    \item The control parameter $y_i$ is chosen such that the size of $G_i$ is approximately the same as that of $G$. 
\end{itemize}

\begin{figure}[!htb] 
	\centering
	\includegraphics[trim={0cm 6.5cm 0 6.5cm}, clip, scale=0.4]{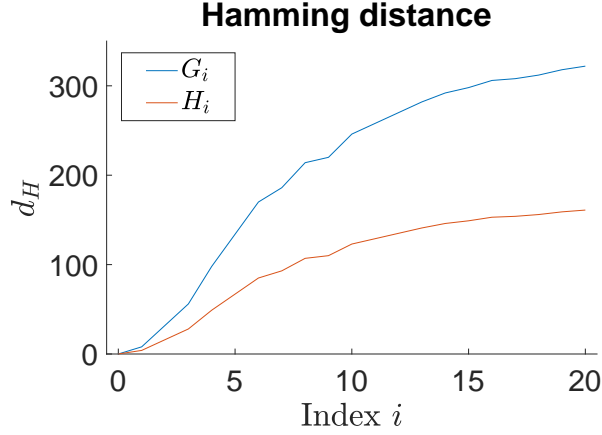}
	\caption{The plot of Hamming distance against the index $i$.}\label{fig:hamming}
\end{figure}

We first study how each $G_i, 1\leq i\leq 20$ is different from $G_0=G$. We compute the Hamming distance $d_H(G_0,G_i)$ that counts the number of edges contained exclusively in either $G_0$ or $G_i$. The plot of $d_H(G,G_i)$ against $i$ is shown in \figref{fig:hamming}. We see that the curve is in general increasing in $i$. It is steep initially and becomes flatter when $i$ is large, say exceeds $10$. For the extreme case $G_{20}$, it is constructed almost solely from sample signals $F$. However, the Hamming distance $d_H(G,G_{20})$ suggests that around $2/3$ of the edges of $G$ are also contained in $G_{20}$, though $G$ and $G_{20}$ are constructed from completely different means. 

\begin{figure}[!htb] 
	\centering
	\includegraphics[trim={0cm 6.5cm 0 6.5cm}, clip, scale=0.35]{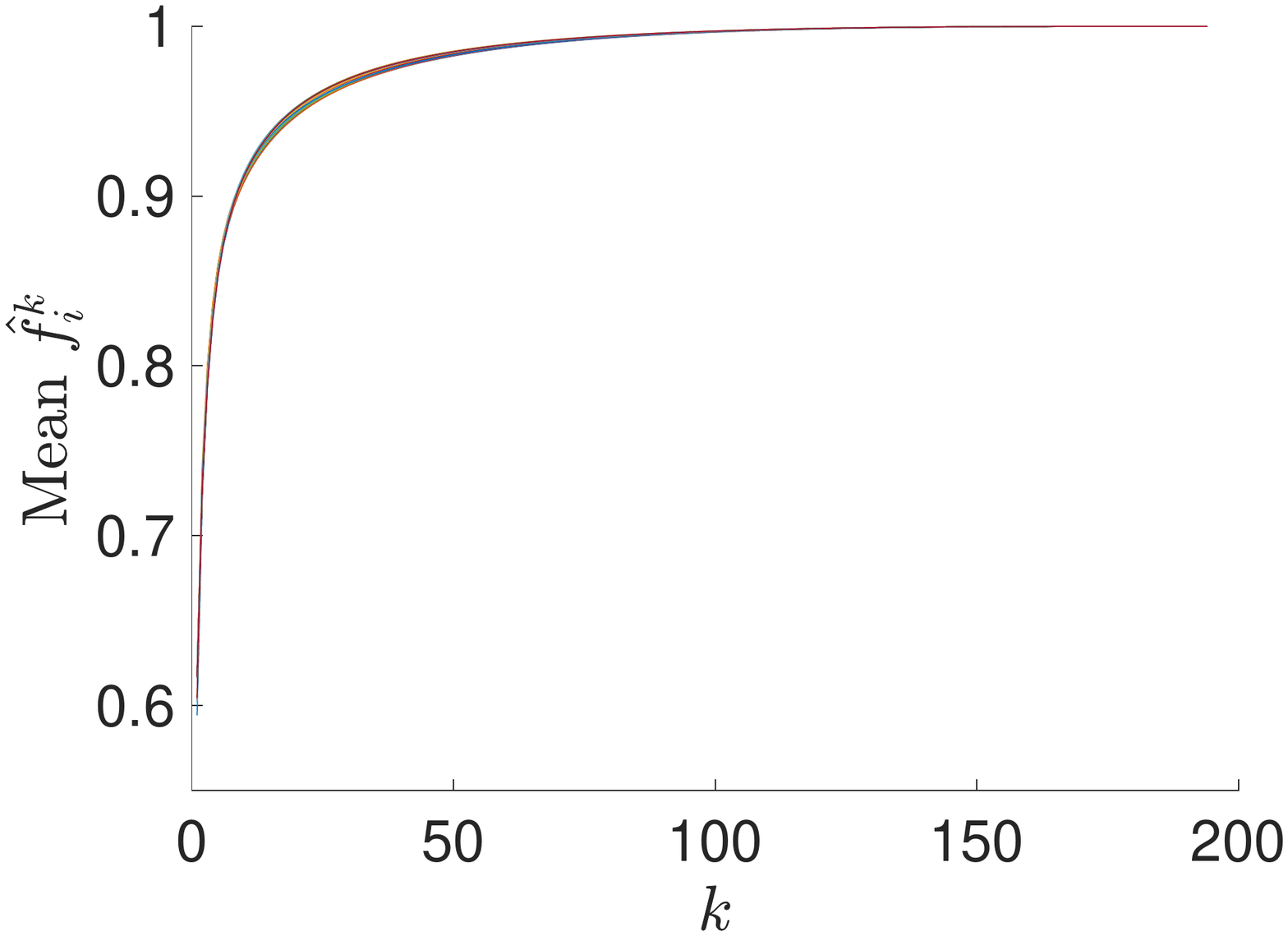}\includegraphics[trim={0cm 6.5cm 0 6.5cm}, clip, scale=0.35]{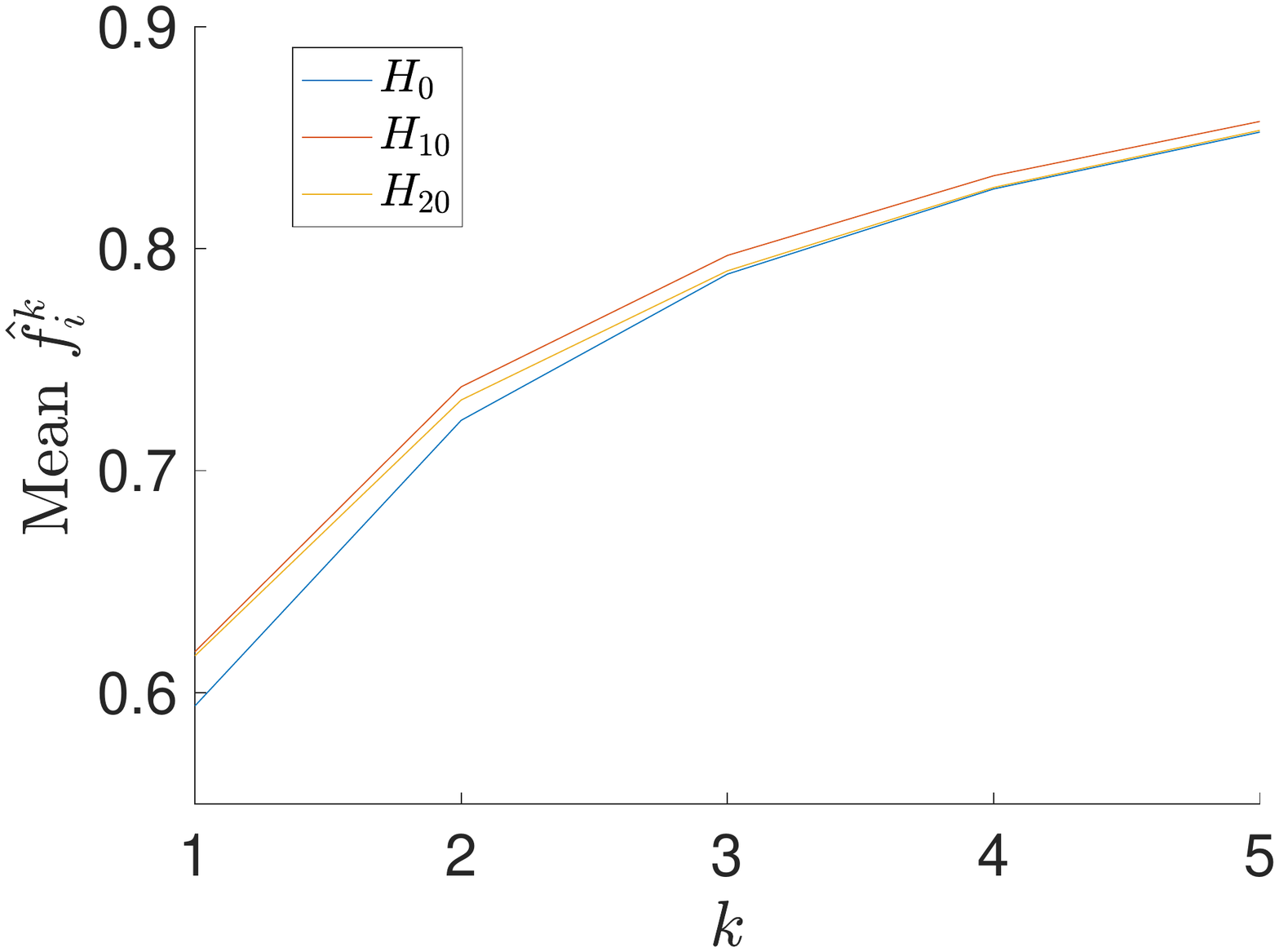}
	\caption{The plots of mean of $\hat{f}_j^k$ against $k$. On the left, we show that general trend by including all $1\leq i\leq 20$ and $1\leq k\leq 194$. On the right, we enlarge the details for $i=0,10,20$ and $k=1,\ldots,5$.}\label{fig:pca12}
\end{figure}

For our next task, we want to perform a Fourier analysis of the temperature signals. It is however unfavorable to use $G_i, 0\leq i\leq 20$ directly as some of them are not connected. We construct $H_i$ whose edges are the unions of edges in $G$ and $G_i$. The plot for $d_{H}(G, H_i)$ is also shown in \figref{fig:hamming}. For example, we add $\approx 25\%, 32\%$ more edges to $G$ to form $H_{10}$ and $H_{20}$ respectively. For each $H_i, 0\leq i\leq 20$, let $L_i$ be its Laplacian and $B_i = \{e_{i,j}, 1\leq j\leq 194\}$ be an eigenbasis of $L_i$. For each signal $f$, we compute its Fourier coefficients $\hat{f}_{i,j} = \langle f, e_{i,j}\rangle$. The principal components of $f$ w.r.t.\ $B_i$ are those indices $j$ such that $|\hat{f}_{i,j}|$ is large. Let $\tau_i$ be the permutation of indices $\{1,\ldots,194\}$ such that $|\hat{f}_{i,\tau_i(j)}|\geq |\hat{f}_{i,\tau_i(j+1)}|$, i.e., $\tau_i$ re-orders the indices according to $|\hat{f}_{i,j}|$ decreasingly. For each $k\leq 194$, we set $\hat{f}^k_i= \sqrt{\sum_{1\leq j\leq k}\hat{f}_{i,\tau_i(j)}^2}/|f|$. If we interpret $|f|$ as the energy of $f$, then $\hat{f}_i^k$ computes the percentage of the energy of $f$ contained in $k$ principal components. Therefore, for the same (small) $k$, the $H_i$ with larger $\hat{f}_i^k$ value is preferred. In \figref{fig:pca12}, we show the plots of average $\hat{f}^k_i$ against $k$. On the left, we show the general trend by including all $1\leq i\leq 20$ and $1\leq k\leq 194$. We see that on the large scale, we have the same general trend for every $H_i,0\leq i\leq 20$. On average, for each $i$, $\hat{f}^k_i$ quickly reaches a very high percentage as $k$ increases, and the curve becomes almost flat. On the other hand, by amplifying the details for $i=0,10,20$ and $k=1,\ldots, 5$, we see that $H_{10}$ gives the largest mean $\hat{f}^j_i$, while $H_0$ gives the smallest. The graph $H_{10}$ uses both information from the graph $G=H_0$ and the sample signals $F$.

\begin{figure}[!htb] 
	\centering
	\includegraphics[trim={0cm 6cm 0 6cm}, clip, scale=0.35]{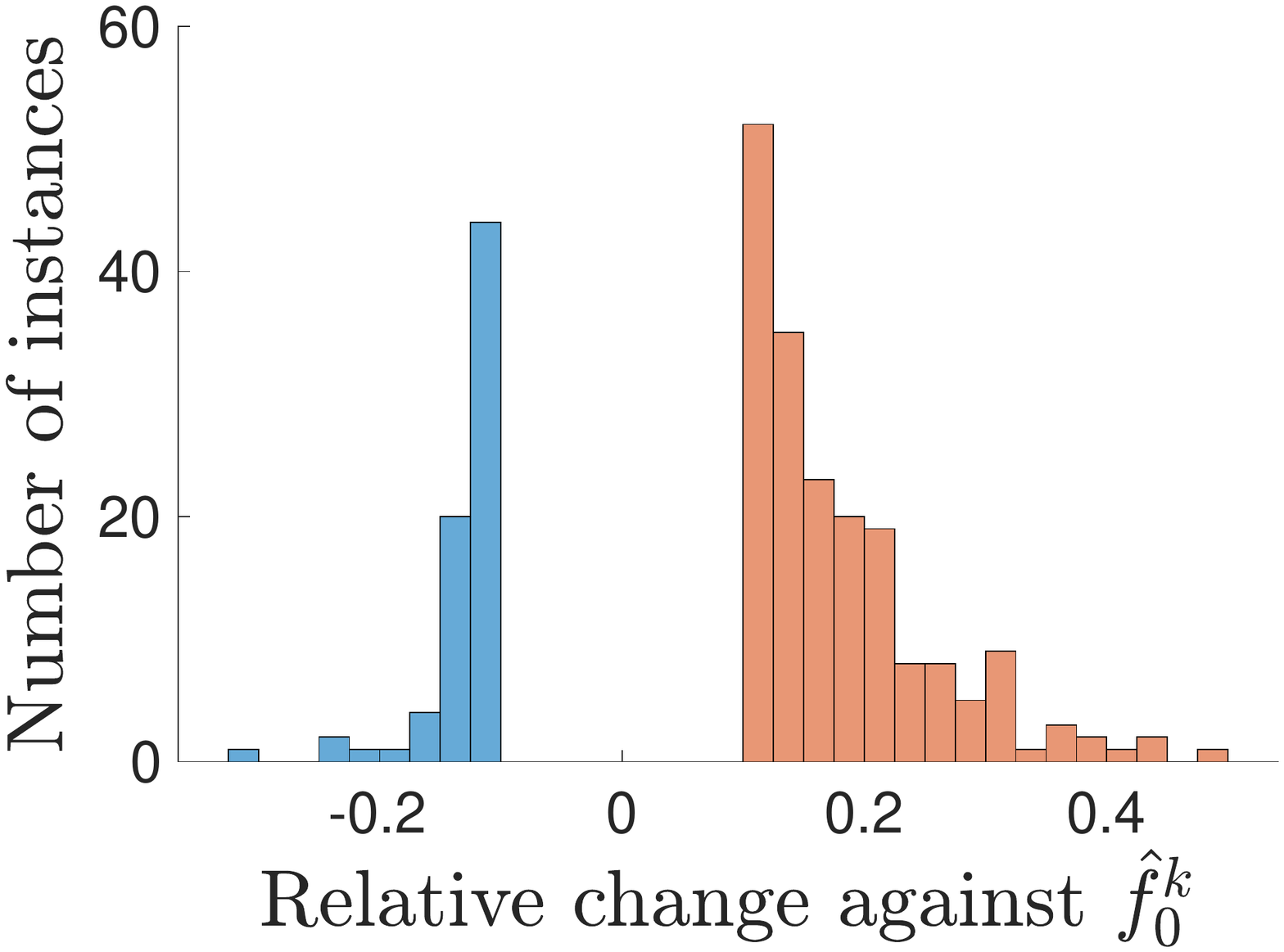}
	\includegraphics[trim={0cm 6cm 0 6cm}, clip, scale=0.35]{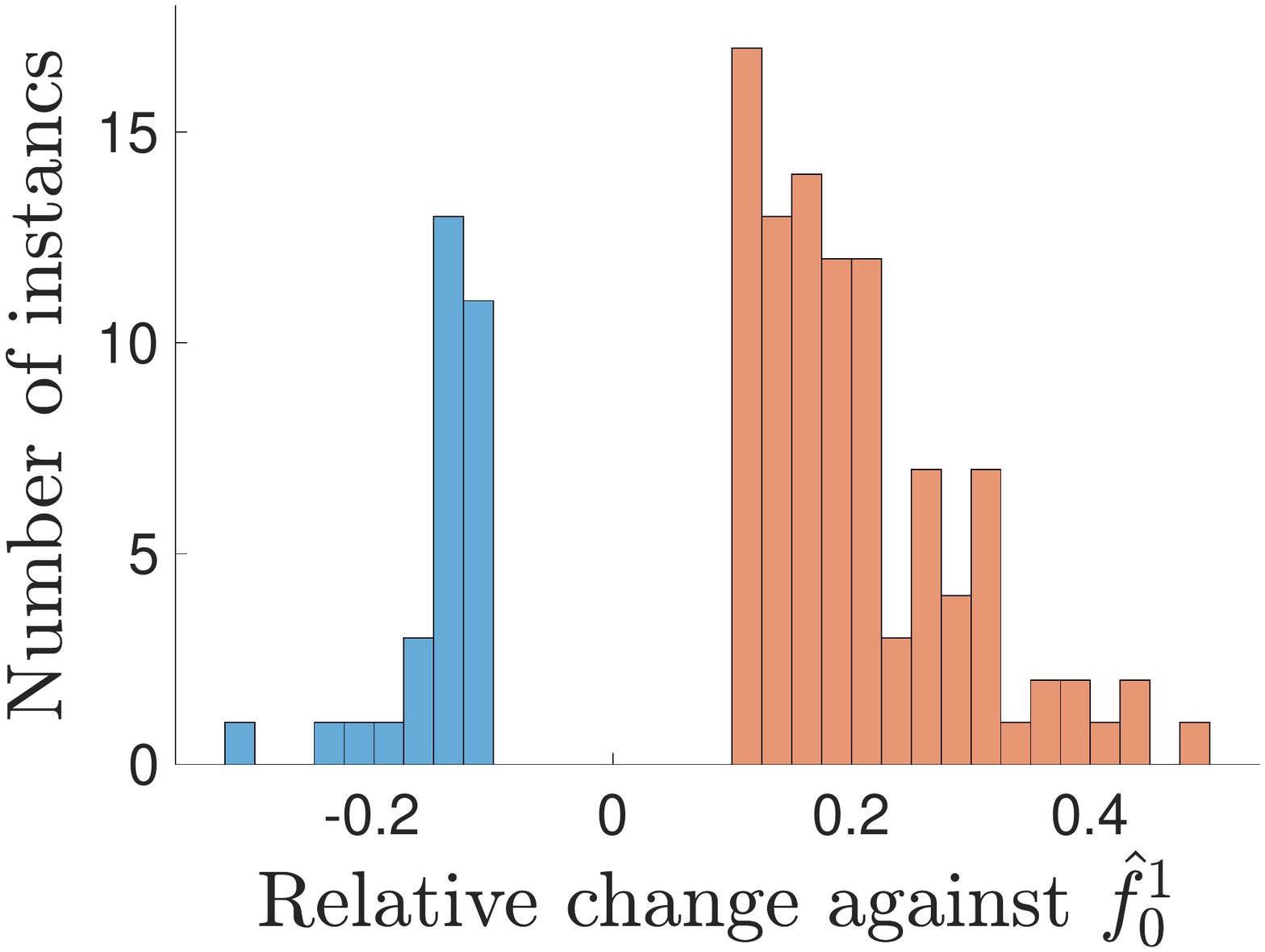}
	\includegraphics[trim={0cm 6cm 0 6cm}, clip, scale=0.35]{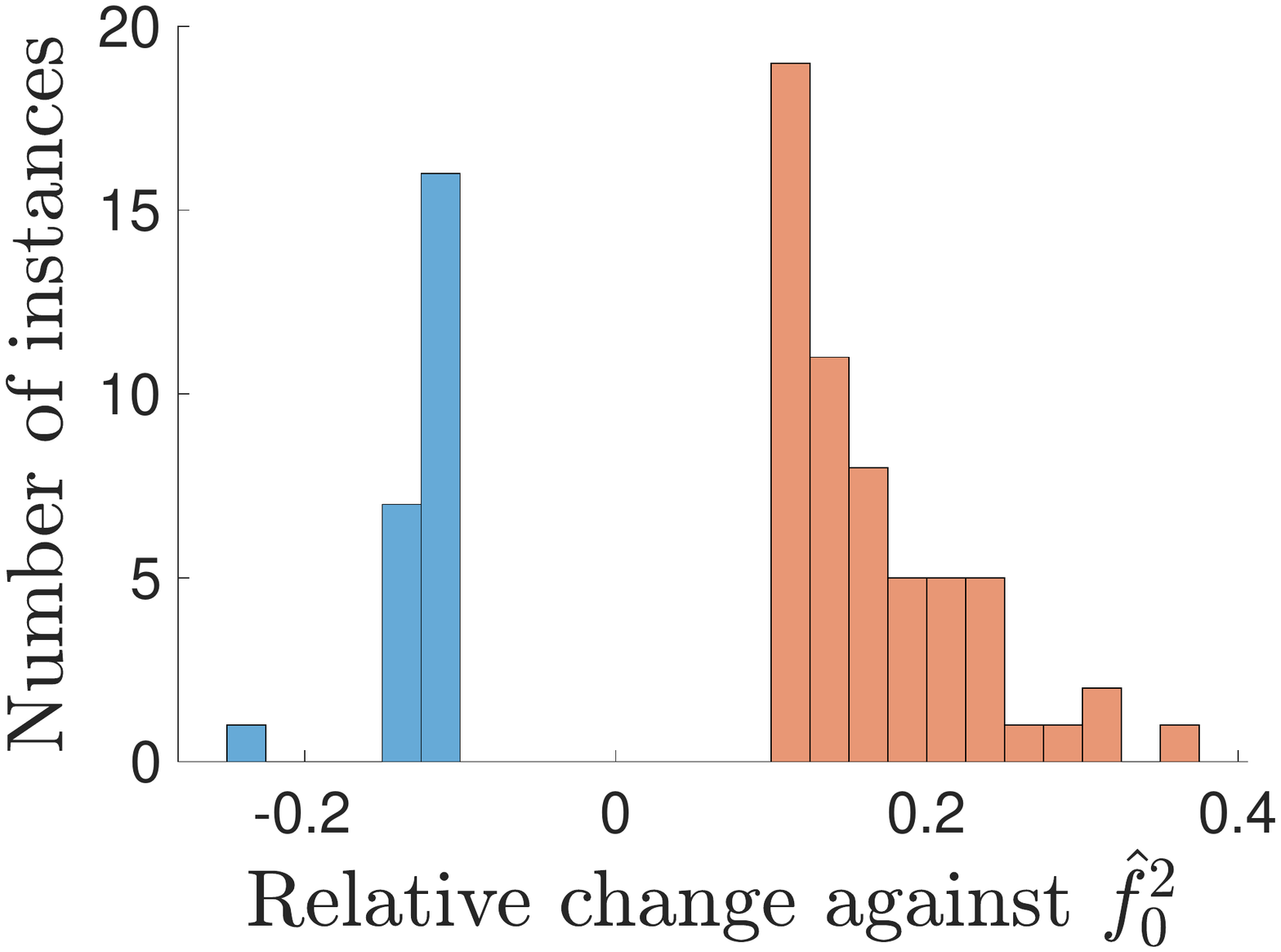}
	\caption{The histograms of distributions of instances for significant relatives changes against $\hat{f}^k_0$. For the first plot, we show the results for $k=1,2,3,4,5$ all considered together. While for the remaining two plots, we show the individual results for $k=1$ and $k=2$.}\label{fig:pca345}
\end{figure}

While \figref{fig:pca12} gives a rough comparison, we now compare $H_0=G$ and $H_{10}$ in more detail. For a signal $f$ and $1\leq k\leq 5$, we find out $(\hat{f}^k_{10}-\hat{f}^k_0)/\hat{f}^k_0$ as the relative change against $\hat{f}^k_0$. We consider the change significant if either $(\hat{f}^k_{10}-\hat{f}^k_0)/\hat{f}^k_0 \geq 10\%$ or $(\hat{f}^k_{10}-\hat{f}^k_0)/\hat{f}^k_0 \leq -10\%$, with the former favors $H_{10}$ and the latter favors $H_0$. We go through every $f$ in the dataset, and the distribution of instances for significant relative changes are shown in \figref{fig:pca345}. We see there are much more instances that favor $H_{10}$ than those that favor $H_{0}$. The difference in such instances is more than $15\%$ of the total number of signals for $k=1$, where we use a single component to approximate a signal. It is approximately $10\%$ for $k=2$, where we use two components to approximate a signal.

\section{Conclusions} \label{sec:con}
In this paper, we study graph signal processing by focusing on graph signals themselves. Motivated by the necessity to understand graph signals geometrically, we introduce a new notion of smoothness. It is based on comparing a signal with a graph with both considered elements of an enlarged set of objects. The new notion allows us to partially classify graph signals, and obtain new insights in conjunction with classical GSP tools. For future works, we shall explore the potential of the framework in more application scenarios.   

\appendices

\section{Proofs of theoretical results}\label[Appendix]{sec:pro}

\begin{IEEEproof}[Proof of \cref{lem:fns}]
By re-arranging indices if necessary, we assume that $f_i\leq f_{i+1}$ and $g_i\leq g_{i+1}$ for $1\leq i\leq n-1$. Moreover, as $g$ is orthogonal to the constant vector, $f$ is not orthogonal to $g$ if and only if $f+c$ is not orthogonal to $g$. Without loss of generality, by adding a constant $c$ to $f$ if necessary, we assume that $g_i\leq f_i\leq g_{i+1}, 1\leq i\leq n-1$ and $f_n\geq g_n$.

As $g$ is nonzero and $g$ is orthogonal to the constant vectors, we have $g_1<0$ and $g_n>0$. Suppose $f_i$ and $g_i$ have the same parity, i.e., $f_ig_i\geq 0$, for each $1\leq i\leq n$. As $f$ is nonzero, either $f_1<0$ or $f_n>0$. Then $\langle f,g\rangle = \sum_{1\leq i\leq n}f_ig_i \geq \max\{f_ng_n, f_1g_1\} >0$. 

If for some $1\leq i\leq n$, $f_ig_i<0$, we have $g_i<0<f_i$ by the interlacing property. For $j>i$, we have $f_j\geq g_j\geq f_i>0$; while for $l<i$, we have $g_l\leq f_l \leq g_i<0$. In particular, $f_jg_j>0$ for each $j\neq i$. Therefore, if $1<i<n$, we find $\langle f, g\rangle \geq f_1g_1+f_ig_i+f_ng_n$. As either $f_1g_1+f_ig_i \geq 0$ or $f_ig_i+f_ng_n\geq 0$, we have $\langle f, g\rangle>0$. If $i=1$ or $i=n$, then $\langle f, g\rangle \geq f_1g_1+f_2g_2+f_ng_n \geq f_2g_2>0$. This concludes the proof. 
\end{IEEEproof}

\begin{IEEEproof}[Proof of \cref{prop:agw}]
We first verify the recoverability of graphs. If $G_F$ is as constructed, then $G$ is nothing but $G_F(1,0\ldots,0)$. For recoverability of signal, we notice that 
\begin{align*}
|f(v_i)-f(v_j)| = \inf_{x_0 \in \mathbb{R}^{\geq 0}} \{G_f(x_0,1) \text{ contains } (v_i,v_j) \text{ as an edge}\}.
\end{align*}
Therefore, if $G_f = G_{f'}$, then $|f(v_i)-f(v_j)| = |f'(v_i)-f'(v_j)|$. In the reverse direction, if $f=af'+c$, then notice that $|f(v_i)-f(v_j)| = |a||f'(v_i)-f'(v_j)|$. The equivalence between $M_f$ and $M_{f'}$ is given by $\phi: M_f \to M_{f'}, (x_0,x_1) \mapsto (x_0,x_1/|a|)$, and $\psi: M_{f'}\to M_f, (x_0,x_1) \mapsto (x_0, |a|x_1)$.

For composability, consider two finite sets of signals $F_1$ of size $k_1$ and $F_2 = \{f_1,\ldots, f_{k_2}\}$ of size $k_2$. Let $F=F_1\cup F_2$ of size $k$. Define $M_{1,2}$ (as an abbreviation of $(M_{F_1})_{F_2}$) to be $\mathbb{R}_{\geq 0}^{k_1+k_2+1}$. An element of $M_{1,2}$ is denoted by $x=(x_0,y_1,\ldots, y_{k_1},z_1,\ldots, z_{k_2})$ and abbreviated by $(x_0,y,z)$ with $y=(y_1,\ldots,y_{k_1})$ and $z = (z_1,\ldots, z_{k_2})$. We define $G_{1,2}:M_{1,2} \to \mathcal{G}$. Given $x = (x_0,y,z)$, $G_{1,2}(x)$ is the following graph. For a pair of nodes $v_i,v_j$, similar to above, let 
\begin{align*}
D_y(v_i,v_j) = \Big(\inf_{x_0 \in \mathbb{R}^{\geq 0}} \{G_{F_1}(x_0,y) \text{ contains } (v_i,v_j) \text{ as an edge}\}\Big).
\end{align*}
From this, we compute 
\begin{align*}
D_{y,z}(v_i,v_j)^2 = D_y(v_i,v_j)^2 + \sum_{1\leq l\leq k_2}z_l(f_l(v_i)-f_l(v_j))^2.
\end{align*}
The nodes $v_i,v_j$ are connected by an edge in $G_{1,2}(x)$ if $D_{y,z}(v_i,v_j) \leq x_0$.

To show that $G_{1,2} \sim G_{F}$, recall an element of $M_{1,2}$ takes the form $(x_0,y_1,\ldots, y_{k_1}, z_1,\ldots, z_{k_2})$ and an element of $M_{F}$ takes the form $(x_0,x_1,\ldots,x_k)$. We examine each signal in $F_1\cup F_2$. For $f\in F_1\cup F_2$, assume it corresponds to $y_i$ if $f\in F_1$, $z_j$ if $f\in F_2$, and $x_l$. Then define $\phi: M_{1,2} \to M_F$ such that the $x_l$-component of $\phi\big((x_0,y_1,\ldots, y_{k_1},z_1,\ldots,z_{k_2})\big)$ is $y_i+z_j$. It is straightforward to check that any smooth map such that $\phi\circ \psi = Id_{M_F}$ defines an equivalence between $G_{1,2}$ and $G_{F}$.
\end{IEEEproof}

\begin{IEEEproof}[Proof of \cref{prop:imt}]
We define the maps $\tau_1$ - $\tau_5$ in \figref{fig:18}.
\begin{itemize}
    \item For $\tau_1$, consider $G=(V,E) \in \mathcal{MG}_n$, define $\tau_1(G)$ to be the metric space on $V$ with distance metric $d_G$ on $G$. 
    \item For $\tau_2$, given $(V,d)$ a finite metric space on $n$ points with metric $d$, define $M_V = \mathbb{R}^{\geq 0}$. For $x_0\in M_V$, $\tau_2(V)(x_0)$ is the graph on $V$ such that distinct pair $v_i,v_j \in V$ are connected by an edge if $d(v_i,v_j) \leq x_0$.  
    \item For $\tau_3$, given $\phi: M \to \mathcal{G}_n$, the associated sequence is $Im(\phi)$ ordered by the number of edges.
    \item For $\tau_4$, let $S=\{G_1,\ldots, G_k\}$ be a finite sequence of graphs on $n$ vertices. If $k=1$, then $\tau_4(S) = G_1$; and if $k\geq 2$, then $\tau_4(S)= G_2$. 
    \item For $\tau_5$, let $f=(f_i)_{1\leq i\leq n}$ (the notation is only valid in this proof) be an element of $\mathcal{F}_n$. Define $M_f = \mathbb{R}_{\geq 0}$. For $x_0\in M_f$, $\tau_5(f)(x_0)$ is the graph on $\{v_1,\ldots, v_n\}$, and $v_i,v_j$ is connected by an edge if $|f_i-f_j|\leq x_0$.
\end{itemize}

The claims $\tau_4\circ \tau_3 \circ \tau_2 \circ \tau_1 = Id_{\mathcal{G}_n}$ $\tau_2\circ\tau_1(G) = G_c$ are straightforward to check for the constructions, which are omitted here. Let us verify the statement on $\tau_5$.    

As we have seen earlier, $\tau_5(f) = \tau_5(g)$ implies that $|f_i-f_j| = |g_i-g_j|$ for any $1\leq i\neq j\leq n$. Let us briefly recall the reason is that 
\begin{align*}
|f_i-f_j| = \inf_{x_0 \in M_f} \{\tau(f)(x_0) \text{ contains } (v_i,v_j) \text{ as an edge}\}.
\end{align*}
Subtracting $f$ and $g$ by the constant signals $f_0$ and $g_0$ respectively, we assume that $f_0=g_0=0$. If $f$ and $g$ are constant signals, then we are done. Otherwise, without loss of generality, we assume that $f_2\neq 0$ and $g_2\neq 0$. If $f_1=0$ and $f_2\neq 0$ are both fixed, then knowing $|f_i-f_1|$ and $|f_i-f_2|$ uniquely determines $f_i$. Therefore, as $|f_i-f_j| = |g_i-g_j|$ for any $1\leq i\neq j\leq n$, we see that $f=g$ if $f_2=g_2$ and $f=-g$ if $f_2=-g_2$.  
\end{IEEEproof}

\begin{IEEEproof}[Proof of \cref{lem:ifi}]
\begin{enumerate}[1)]
    \item We notice that $*_{f+c} = *_f$ and $*_{rf} = r(*_f)$ (as maps). Therefore, if $f$ is $\epsilon$-smooth, then so do $rf$ and $f+c$.
    \item Following directly from the definition, a signal of $f$ is $0$-smooth if and only if $Im(*_f) \subset Im(G_c)$. Suppose for $0$-smooth signal $f$, there is an edge $(v_i,v_j)$ such that $f(v_i)=f(v_j)$. If $f$ is not a constant signal, then we can always find another edge $(v_k,v_l)$ such that $f(v_k)\neq f(v_l)$ (as $G$ is connected). Choose $0<x< |f(v_k)-f(v_l)|$, then $*_f(x)$ contains $(v_i,v_j)$ and does not contain the edge $(v_k,v_l)$. Hence, $*_f(x) \notin Im(G_c)$, and this contradicts that $f$ is $0$-smooth.
    
    Suppose $G$ is a path graph on $n\geq 2$ vertices. We order the vertices as $v_1,\ldots, v_n$ from one end to the other. Then the non-constant signal $f=(1,\ldots, n)^T$ is also $0$-smooth.
    
    Conversely, suppose $G$ is not a path graph and $f$ is a non-constant $0$-smooth signal. According to the first paragraph, for each edge $(v_i,v_j)$ of $G$, $f(v_i)\neq f(v_j)$. We consider two cases, $G$ is a tree and $G$ contains a cycle. 
    
    Case $1$: If $G$ is a tree, then it must contain a node of degree at least $3$, say $v_1$. Without loss of generality, assume that $3$ of its neighbors are $v_2$, $v_3$, and $v_4$. If $|f(v_i)-f(v_1)|, i=2,3,4$ are not the same, then re-ordering if necessary we may assume $|f(v_2)-f(v_1)|<x< |f(v_3)-f(v_1)|$ for some $x$. Then $*_f(x)$ contains $(v_1,v_2)$ as an edge and excludes $(v_1,v_3)$. Hence, $*_f(x)\notin Im(G_c)$ and this is a contradiction. If $|f(v_i)-f(v_1)|,i=2,3,4$ are all equal to $y$, then for at least two nodes, say $v_2,v_3$, we have $f(v_2)=f(v_3)$. Choose $0<x<y$, we have $*_f(x)$ contains $(v_2,v_3)$ as an edge but excludes both $(v_1,v_2)$ and $(v_2,v_3)$. Hence, $*_f(x) \notin Im(G_c)$ and this leads to a contradiction.
    
    Case $2$: If $G$ contains a cycle of size $m \geq 3$. We order the vertices along the cycle as $v_1,\ldots,v_m$. If there are pairs $(v_i,v_{i+1})$ and $(v_j,v_{j+1})$ with $i\neq j$ (by convention, $v_{m+1}=v_1$), such that $|f(v_i)-f(v_{i+1}))|<x<|f(v_j)-f(v_{j+1})|$, then $*_f(x)$ contains $(v_i,v_{i+1})$ as an edge and excludes $(v_j,v_{j+1})$. This is impossible if we want $*_f(x) \notin Im(G_c)$. Therefore, $|f(v_i)-f(v_{i+1})|$ equals to some $y\neq 0$ for every $1\leq i\leq m$. This can only happen that for some $i\neq j$, $f(v_i)=f(v_j)$. Consider $0<x<y$, $*_f(x)$ contains $(v_i,v_j)$ as an edge but excludes $(v_1,v_2)$. Hence $*_f(x) \notin Im(G_c)$ and the contradiction concludes the last subcase. 
    \item Notice that in the expression (\ref{eq:dms}) of $d_{\mathcal{S}}$ with $\gamma_1 = *_f$ and $\gamma_2 = G_c$, the $d_H$ factor in each term is uniformly bounded by $2n^2$ independent of $f$. Moreover, fix a non-constant signal $f$, for any signal $f'$ such that $\norm{f-f'}\leq \epsilon$ for $\epsilon$ small enough, we can always ensure that 
    \begin{enumerate}[(a)]
        \item $f'$ is non-constant.
        \item The part $1/|\gamma_1'^{-1}(Im(\gamma_1')^{\circ})|$ has an upper bound $\alpha$ that depends only on $f$ for any $\epsilon$ small enough. 
        \item For any $G\in Im(\gamma_1)^{\circ} \cup Im(\gamma_1')^{\circ}$, $\big| |\gamma_1^{-1}(G)|-|\gamma_1'^{-1}(G)|\big|$ is bounded by $\beta(\epsilon)$ that converges to $0$ if $\epsilon\to 0$.
    \end{enumerate}
    We also notice that the number of graphs in $Im(\gamma_1)^{\circ} \cup Im(\gamma_1')^{\circ}$ is also bounded by $n^2$, which is independent of both $f$ and $\epsilon$. Now we estimate 
    \begin{align*}
    |d_{\mathcal{S}}(\gamma_1,\gamma_2)-d_{\mathcal{S}}(\gamma_1',\gamma_2)| \leq n^2\cdot\beta(\epsilon)\cdot 2\alpha\cdot 2n^2=4n^4\alpha\beta(\epsilon) \to 0,
    \end{align*}
    as $\epsilon \to 0$.
\end{enumerate}
\end{IEEEproof}

\begin{IEEEproof}[Proof of \cref{prop:leb}]
Let $\gamma_1 = *_f$, $\gamma_2=G_c$ and ordered sequence of pairs of vertices $Q = (Q_l)_{1\leq l\leq n(n-1)/2}$ as earlier. As we assume that $|f(u_1)-f(v_1)|\neq |f(u_2)-f(u_2)|$, the image $Im(\gamma_1)$ of $\gamma_1$ is a sequence of graphs $G_{(l)}, 1\leq l\leq n(n-1)/2$ such that the edge set of $G_{(l)}$ is $\{Q_l, 1\leq l\leq l\}$, i.e., $G_{(l+1)}$ is obtained from $G_{(l)}$ by adding $Q_{l+1}$. To compute $\epsilon$, we need to find $d_H(G_{(i)},G^{(j)})$ for suitable $1\leq i\leq n(n-1)/2$ and $0\leq j\leq k$. In doing so, we may take the sum over all the edges of $G_{(i)}$ and $G^{(j)}$, i.e., $d_H(G_{(i)}, G^{(j)}) = \sum_{1\leq l \leq n(n-1)/2} \theta_{i,j}(Q_l)$, where $\theta_{i,j}(Q_l) = 1$ if $Q_l$ is an edge of either $G_{(i)}$ or $G^{(j)}$ exclusively and $0$ otherwise. Therefore, $\epsilon$ is expressed in the form $\sum_{G_{(i)},G^{(j)}}\sum_{Q_l} \theta_{i,j}(Q_l)$. We want to change the summation order  $\sum_{Q_l}\sum_{G_{(i)},G^{(j)}} \theta_{i,j}(Q_l)$. This prompts us to go through $Q_l$ one-by-one and study its membership in $G_{(i)}$ and $G^{(j)}$. With this perspective, we describe an equivalent formula for $\epsilon_{\mathcal{P}}$.

For the partition $\mathcal{P}$ of $I$, we claim that \begin{align} \label{eq:ess}
    \epsilon_{\mathcal{P}} = \sum_{0\leq j\leq D_G}\sum_{i \in I_j} \frac{\gamma_1^{-1}(G_{(i)})}{\gamma_1^{-1}(Im(\gamma_1)^{\circ})} d_H(G_{(i)}, G^{(j)}).
\end{align}
We prove the claim by showing that for each pair of nodes $Q_l = \{u_l,v_l\}, 1\leq l \leq n(n-1)/2$, it contributes to the same summand on both sides of (\ref{eq:ess}). For the left-hand-side, the contribution of $Q_l$ is $\epsilon_l$. On the right-hand-side, $Q_l$ contributes to either $1$ or $0$ in each of $d_H(G_{(i)},G^{(j)})$. We notice that $G_{(i)}$ is a subgraph of $G_{(i+1)}$ and similarly $G^{(j)}$ is a subgraph of $G^{(j+1)}$. The pair $Q_l$ is an edge of $G_{(i)}$ for $i\geq l$. Let $k$ be the smallest index such that $Q_l$ is an edge of $G^{(k)}$. Then $Q_l$ is an edge of $G^{(j)}$ for $j\geq k$. Therefore, $Q_l$ contributes $1$ to $d_{H}(G_{(i)}, G^{(j)})$ with $i\in I_j$ if and only if:
\begin{enumerate}[(a)]
    \item \label{it:ilj} $i < l$ and $j \geq k$, or
    \item \label{it:igl} $i \geq l$ and $j < k$.
\end{enumerate}
Let $l'$ be smallest index such that $Q_{l'} \in G^{(k)}$ and $l \in I_{k'}$. Correspondingly, we consider two cases on $k$ and $k'$ (illustrated in \figref{fig:21}):
\begin{enumerate}[(a')]
    \item If $k\leq k'$, then \ref{it:igl} cannot happen. For if $i\geq l$, then $i\notin I_j$ as $j<k\leq k'$. Therefore $i<l$ and $j\geq k$. The condition $i\in I_j$ together with $j\geq k$ imply that $i\geq l'$. In summary, we have $l'\leq i<l$ and $k\leq j\leq k'$ such that $i\in I_j$. Therefore, the contribution of $Q_l$ to the right-hand-side of (\ref{eq:ess}) is 
    $\frac{\sum_{l'\leq i< l}\gamma_1^{-1}(G_{(i)})}{\gamma_1^{-1}(Im(\gamma_1)^{\circ})}$, which is exactly $\epsilon_l$ as $\gamma_1^{-1}(G_{(i)}) = |f(u_{i+1})-f(v_{i+1})|-|f(u_i)-f(v_i)|$.
    \item If $k>k'$, then \ref{it:ilj} cannot happen. We can show that the contribution of $Q_l$ to the right-hand-side of (\ref{eq:ess}) is $\epsilon_l$ by the exact same argument.
\end{enumerate}
This completes the proof of (\ref{eq:ess}).
\begin{figure}[!htb] 
	\centering
	\includegraphics[scale=0.5]{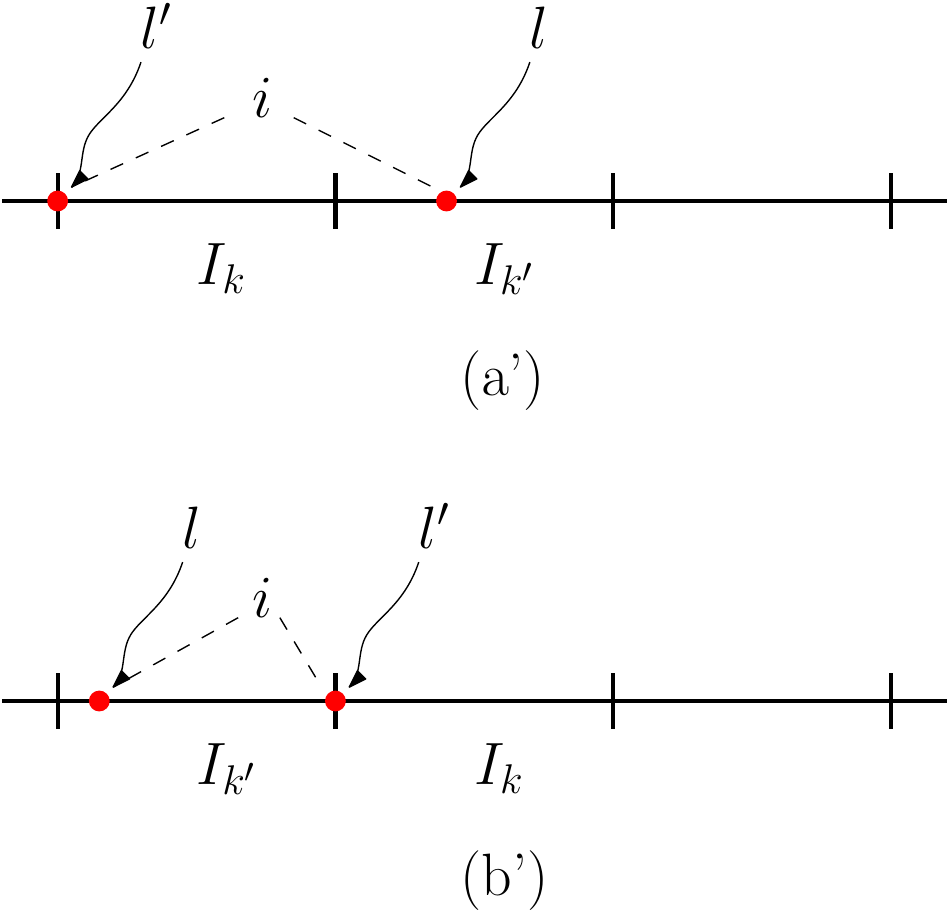}
	\caption{An illustration of the proof of (\ref{eq:ess}).}\label{fig:21}
\end{figure}	

If we examine the definition of $\epsilon$, for each $0\leq j \leq D_G$, define $I_j'$ as follows: an index $1 \leq l \leq n(n-1)/2$ belongs to $I_j'$ if $d_H(G_{(l)}, G^{(j')}) >d_H(G_{(l)}, G^{(j)}) = d_H(G_{(l)}, Im(\gamma_2))$ for any $j'>j$. We claim that $\mathcal{P}' = \{I_j', 0\leq j\leq D_G \}$ is a partition of $Q$. 

For any $1\leq l\leq n(n-1)/2$, suppose $l$ is associated with $j$ and $l+1$ is associated with $j'$ as described in the previous paragraph. To prove the claim, it suffices to show that $j\leq j'$. Suppose on the contrary that $j'<j$. From the definition, we have $d_H(G_{(l)}, G^{(j)}) \leq d_H(G_{(l)}, G^{(j')})$ and $d_H(G_{(l+1)}, G^{(j')}) < d_H(G_{(l+1)}, G^{(j)})$. Notice that $G_{(l+1)}$ is obtained from $G_l$ by including a single edge $Q_{l+1}$. We consider the following cases:
\begin{enumerate}[(i)]
    \item \label{it:qgi} $Q_{l+1} \in G^{(j')}$: In this case, $Q_{l+1}$ is also in $G^{(j)}$. Hence, $d_H(G_{(l)}, G^{(j)})-1 = d_H(G_{(l+1)}, G^{(j)})$ and $d_H(G_{(l)}, G^{(j')})-1 = d_H(G_{(l+1)}, G^{(j')})$. This gives a contradiction, as
    \begin{align*}
     d_H(G_{(l+1)}, G^{(j)})= d_H(G_{(l)}, G^{(j)})-1 \leq d_H(G_{(l)}, G^{(j')})-1 = d_H(G_{(l+1)}, G^{(j')}).
    \end{align*}
    \item $Q_{l+1} \notin  G^{(j)}$: We also have $Q_{l+1} \notin G^{(j')}$. Hence, $d_H(G_{(l)}, G^{(j)})+1 = d_H(G_{(l+1)}, G^{(j)})$ and $d_H(G_{(l)}, G^{(j')})+1 = d_H(G_{(l+1)}, G^{(j')})$. We have the same contradiction as in \ref{it:qgi}.
    \item $Q_{l+1} \in G^{(j)}$ and $Q_{l+1} \notin G^{(j')}$: In this case, we have $d_H(G_{(l)}, G^{(j)})-1 = d_H(G_{(l+1)}, G^{(j)})$ and $d_H(G_{(l)}, G^{(j')})+1 = d_H(G_{(l+1)}, G^{(j')})$. The equations imply that $d_H(G_{(l+1)},G^{(j)})<d_H(G_{(l+1)}, G^{(j)})$, which is again a contradiction.
\end{enumerate}
The three cases conclude our proof of the claim by contradiction.

In defining the partition $\mathcal{P}'$, for the pair $l$ and $I_j'$, we have $d_H(G_{(l)}, G^{(j)}) = d_H(G_{(l)}, Im(\gamma_2))$. Therefore for $\mathcal{P'}$, we have $\epsilon = \epsilon_{\mathcal{P'}}$, in view of (\ref{eq:ess}). For any other partition $\mathcal{P} = \cup_{0\leq k\leq D_G} I_k$, if $i\in I_{j'}$, then $d_H(G_{(l)}, G^{(j')}) \geq d_H(G_{(l)}, Im(\gamma_2)) = d_H(G_{(l)}, G^{(j)})$. Therefore, $\epsilon_{\mathcal{P}} \geq \epsilon_{\mathcal{P'}}$. 
\end{IEEEproof}

\bibliographystyle{IEEEtran}
\bibliography{IEEEabrv,StringDefinitions,allref}

\end{document}